\documentclass[aps,pre,superscriptaddress,showkeys]{revtex4-1}

\usepackage{ae,aecompl}
\usepackage[T1]{fontenc}
\usepackage[utf8x]{inputenc}
\usepackage{verbatim}
\usepackage{amsmath}
\usepackage{graphicx}
\usepackage{amssymb}
\usepackage{latexsym}
\usepackage{amsxtra}
\usepackage{amsfonts}
\usepackage{mathrsfs}%
\usepackage{type1cm}
\usepackage{epsfig}
\usepackage{enumerate}
\usepackage{stmaryrd}
\usepackage{appendix}
\usepackage{mdwlist}
\usepackage{geometry}

\long\def\symbolfootnote[#1]#2{\begingroup%
\def\thefootnote{\fnsymbol{footnote}}\footnote[#1]{#2}\endgroup} 

\usepackage{vmargin}
\usepackage[colorinlistoftodos, shadow]{todonotes}

\newcommand{\CHANGED}[1]{{#1}}

\newcommand{\ADDED}[1]{{#1}}

\newcounter{mycomment}

\DeclareGraphicsExtensions{pdf}
\graphicspath{{fig/}}
\renewcommand{\BibitemShut}[1]{}

\begin{document}


\title{Bootstrapping under constraint for the assessment of group behavior in\\ human contact networks}


\author{Nicolas Tremblay}
\thanks{Corresponding author: firstname.lastname AT ens-lyon DOT fr}
\affiliation{Physics Laboratory, ENS Lyon, Université de Lyon, CNRS UMR 5672, Lyon, France}

\author{Alain Barrat}
\affiliation{Aix Marseille Université, CNRS, CPT, UMR 7332, 13288 Marseille, France}
\affiliation{Université de Toulon, CNRS, CPT, UMR 7332, 83957 La Garde, France}
\affiliation{Data Science Laboratory, Institute for Scientific Interchange (ISI) Foundation, Torino, Italy}

\author{Cary Forest}
\affiliation{University of Wisconsin, Physics Department, Madison, USA}

\author{Mark Nornberg}
\affiliation{University of Wisconsin, Physics Department, Madison, USA}

\author{Jean-Fran\c{c}ois Pinton}
\affiliation{Physics Laboratory, ENS Lyon, Université de Lyon, CNRS UMR 5672, Lyon, France}

\author{Pierre Borgnat}
\affiliation{Physics Laboratory, ENS Lyon, Université de Lyon, CNRS UMR 5672, Lyon, France}



\date{\today}

\begin{abstract}
The increasing availability of time --\;and space\;-- resolved data
describing human activities and interactions gives insights into both
static and dynamic properties of human behavior. In practice,
nevertheless, real-world datasets can often be considered as only one
realisation of a particular event. This highlights a key issue in
social network analysis: the statistical significance of estimated
properties. In this context, we focus here on the assessment of quantitative features 
of specific subset of nodes in empirical networks. We present a method of statistical resampling based
on bootstrapping groups of nodes under constraints within the
empirical network. The method enables us to define acceptance intervals for various
Null Hypotheses concerning relevant properties of the subset of nodes under consideration, in order
to characterize by a statistical test its behavior as ``normal'' or not.
We apply this method to a high resolution dataset describing the
face-to-face proximity of individuals during two co-located
scientific conferences. As a case study, we show how to probe whether
co-locating the two conferences succeeded in bringing
together the two corresponding groups of scientists.
\end{abstract}


\maketitle

\section{Introduction}
\label{sec:1}

High resolution measurements of co-presence and even face-to-face interactions between
individuals in different social gatherings --\;such as scientific
conferences, museums, schools, or hospitals\;-- were made possible in
the recent years by the use of wearable sensors, using bluetooth,
wireless or RFID (Radio Frequency IDentification) technology. These
new data paved the way to many empirical
investigations~\cite{cattuto2010dynamics,eagle2006reality,hui2005pocket,salathe2010high}
of human contacts, both from a static (e.g., existence of communities,
clustering, heterogeneities in the number of contacts...) and dynamic (distribution of the
durations of contacts, of the time between contacts, or of the lifetime of groups of different
sizes...) points of view.

A major issue regarding the analysis of these datasets is that each one of them 
represents a single realisation of a particular event: in contrast
to the study of ensembles of random networks, it is not possible to
generate multiple realizations of the event. Associating a
statistical significance to any measured property of these datasets is
thus a challenging issue. 
\CHANGED{The present work seeks to attribute statistical 
significance, in the form of a statistical test and acceptance intervals, 
to observable features in a network. 
More precisely, the features under study will 
characterize a specific group of nodes within the graph, and we discuss how to address the
question whether this group of nodes is normal or not as compared to
other groups in the network.}

Two data-driven methods have been widely used in the general case to
obtain acceptance intervals for observable features: the jackknife and
bootstrapping~\cite{efron1982jackknife,zoubir2004bootstrap}. Both are
based on drawing random samples from the unique original data recorded
in an observation. Transposing the classical bootstrap approach to the case of data represented by
graphs is however not straightforward. 
\CHANGED{Only a few works have considered resampling of graphs,
for instance via the generation of resampled versions 
of the empirical graph as a whole~\cite{eldardiry2008resampling,ying2009graph}.
Classically, graph resampling methods aim at studying the significance of the
empirical graph structure and topology, for instance for phylogenetic trees~\cite{drummond2007beast}
or bayesian-induced networks~\cite{friedman1999data}. Another
application concerns the significance of community
structures~\cite{Fortunato2010,Lancichinetti2010,rosvall2010mapping} in networks. 
Here, in contrast with these works, we do not perform graph resampling as we consider the whole network 
as a fixed input:
our aim is to use statistical resampling techniques
to design a statistical test and acceptance intervals for observables pertaining to groups of nodes in a 
given (empirical) network.}

In this paper, focusing on features of groups of nodes, we
formulate a bootstrap protocol suited to complex networks. 
To this aim, we focus on a specific group of nodes in the graph,
consider resampled versions of this group of interest within the graph, 
and compare the studied group with its resampled versions. 
A key point is that the resampled groups have to take into account some
dependences or constraints existing in the original group in order to constitute a relevant
bootstrap ensemble. 
We then choose specific group features and compare 
these features in the original group and in the resampled groups. 
This procedure provides a measure of the statistical significance of the chosen features
in the graph.
The developed method allows us to estimate whether a feature
deviates from a normal behavior of this feature in the bootstrap ensemble
(i.e., a Null Hypothesis of a statistical test for this feature).
By combining several features, it enables us to define 
normal behaviors of groups of nodes in the graph and to assess whether
the specifically studied group's behavior is normal or anomalous, and in which respect.

The paper is structured in the following way.  We introduce in
Section~\ref{sec:2} the bootstrapping of groups of nodes in complex
networks and how to use this procedure to devise statistical tests;
this part represents the methodological contribution of this work. 
Then, in order  to illustrate the possibilities offered by the proposed
method for real data of complex networks, we consider a dataset
describing the face-to-face interactions of individuals collected in
two co-located conferences of the American Physical Society, involving
two distinct scientific groups: the Division of Plasma Physics Meeting
and the Gaseous Electronics Conference. We show 
in  Section~\ref{sec:5} how
the proposed method assesses to what extent both groups mix together
during these conferences. 
A conclusion is given in Section~\ref{sec:6}.
Details on the data set as
well as on a validation of our method on controlled benchmarks are
provided in the Appendices.

\section{Bootstrapping and statistical test for complex networks}
\label{sec:2}

Our main objective is to provide statistical significance for measured features associated
to subsets of nodes ('groups') in networks. 
A standard way is to formulate a Null Hypothesis
for the normal behavior of a group, and to perform a statistical test to decide whether or not this Null Hypothesis has to be rejected. 
In this section, we will propose in the context of weighted networks a specific resampling method based on bootstrapping 
constrained groups in the network, and a way to perform statistical tests using this bootstrapping method.

Bootstrapping~\cite{efron1982jackknife,zoubir2004bootstrap} is a well-known data-driven method 
that creates new random pseudosamples by using only one empirical observation of the data. 
\CHANGED{In order to adapt the bootstrapping methodology to our goal, we propose here a specific resampling method to draw replicates of
groups with relevant properties. The method is based on two main ingredients: 
1)~to describe normal behavior, a Null Hypothesis is defined that
imposes constraints on the groups and we propose a computational scheme to draw 
groups that correspond to the proposed Null Hypothesis; 
2)~we build a bootstrap set of many such constrained groups, by randomly sampling them 
independently and with replacement, as in classical bootstrapping.}
Combining these two steps, we are then able to propose a bootstrap test to decide whether the specific group
of interest is compatible with the proposed Null Hypothesis. 
We detail the method in the next paragraphs.

The main advantage of using a bootstrap-inspired technique is
that it does neither require any additional information with respect to the
network itself, \ADDED{nor any model of the network's properties}: it is fully data-driven. Moreover, unlike other data-driven
resampling methods such as the jackknife, it is possible to adjust the
size of the drawn samples to the size of the studied group.

\CHANGED{However, a major and tricky issue arises when assessing significance of some 
features of a specific group in a network: generally, neither the nodes nor the links
are independent from each other. Simple use of descriptive statistics of classical bootstraps
are not suited to deal with dependent data. 
A key point of our work is to propose a protocol suited
to networks, that takes into account some of the dependences in the data.
}

\subsection{Relevant observable features for groups in complex networks}

Let $\mathcal{G} = (\mathcal{V},\mathcal{E})$ be the graph representation of the studied complex network,
with $\mathcal{V}$ its set of nodes and $\mathcal{E}$ its set of edges. 
We call $X^{0} \subset \mathcal{V}$ the chosen subset of nodes whose behavior we want to compare to the 
behavior of ``normal'' groups,
obtained as random bootstrap samples satisfying given constraints as explained above.
Let us call $R^{0} \subset \mathcal{V}$ the remaining nodes of the network 
($R^{0} = \mathcal{V} \backslash X^{0}$).

We quantify $X^{0}$'s ``behavior'' by looking at several observable features that are representative of 
how the group is structured. 
In the context of social networks, relevant features are for instance ones that quantify whether there 
are strong 
contacts inside the group, possibly stronger than with other nodes. 
For the use of the method in Section~\ref{sec:5}, the following observable
features are used (generically referred to as $Z^0$ in the following), in addition to the 
cardinality $M$ of the group $X^{0}$: 
\begin{itemize}
\item $N_{XX}^{0}$ the total number of links of $\mathcal{E}$ between nodes of $X^{0}$;
\item $N_{RR}^{0}$ the total number of links of $\mathcal{E}$ between nodes of $R^{0}$; 
\item $N_{XR}^{0}$ the total number of links of $\mathcal{E}$ connecting the two groups of nodes;
\item $T_{XX}^{0}$ the total weight of the links of $\mathcal{E}$ between nodes of $X^{0}$;
\item $T_{RR}^{0}$ the total weight of the links of $\mathcal{E}$ between nodes of $R^0$;
\item $T_{XR}^{0}$ the total weight of the links connecting the two groups.
\item $Q_X^{0}$ the modularity computed when partitioning the nodes of $\mathcal{G}$ in two groups $X^{0}$ and $R^{0}$. 
\end{itemize}
For completeness, we recall that the modularity of a partition of $\mathcal{G}$ is defined
by~\cite{newman2004analysis}: 
$Q=\frac{1}{2N}\sum_{i \in \mathcal{V}, j \in \mathcal{V}} \left[A_{ij}-\frac{s_{i}s_{j}}{2N}\right]\delta(c_{i},c_{j})$, 
where $A$ is  the weighted adjacency matrix of the graph $\mathcal{G}$,
$s_{i}=\sum_{j\in\mathcal{V}}A_{ij}$ is the strength of node $i$,
$N=\frac{1}{2}\sum_{i \in \mathcal{V}, j \in \mathcal{V}} A_{ij}$ is the sum of all weights, 
and $c_i$ is the label of the group of node $i$, 
so that $\delta(c_{i},c_{j})=1$ if nodes $i$
and $j$ are in the same group, and $0$ otherwise. In the present case of a partition in two groups,
the modularity takes values between $-0.5$ and $0.5$ and measures how well the partition separates 
the network into distinct
communities (a value close to $0.5$ denotes two strong communities) \footnote{
\ADDED{The modularity is lower than $0.5$ when there are two groups because 
the modularity of a partition in $K$ groups is known to be bounded by $1-1/K$~\cite{mieghem2011graph}.}}.

\ADDED{We thus consider overall $F = 7$ observables features (in addition to the cardinality of the group).}
The chosen observables are not fully independent and one might question why we consider so many. 
In particular, one of the 
most widely used observables regarding the behavior of a group in a network is the 
modularity~\cite{newman2004analysis}, and one might
argue that is is enough to consider cardinality and modularity; 
however, modularity is neither a sufficient nor a unique way to discriminate between 
different types of behaviors. Moreover, the studied groups (i.e., of which one wants to know if they are normal or not) 
might not necessarily form communities in the network.
By adding the six other observables that are admittedly not totally independent from the modularity, 
we accept some level of redundancy in the information we gather in order
to yield a more complete and discriminative description of groups.
 
Depending on the specific issue addressed and of the nature of the complex network at hand, 
other observable features could be considered as relevant to describe the behavior of a group. 
We are here
guided by the case study we will consider later, consisting in networks of face-to-face 
contacts between individuals (details
on the data are given in the next Section and in Appendix~\ref{sec:4}), but 
we emphasize that the proposed procedure of bootstrap under constraints is directly usable 
in other contexts. 
 
 \subsection{Bootstrapping protocol for statistical testing of a group in a network}
 \label{subsec:protocol}

Once the $F$ specific features $Z$ are chosen ($Z$ is used as a generic notation for 
any one feature, while $Z^0$ is
the value taken by this feature for the group of interest), the steps 
forming the backbone of the bootstrapping procedure to test the 
normality of a group in a network are as follows:
\begin{enumerate}
\item First, we formulate a Null Hypothesis regarding what is supposed to be 
a ``normal'' behavior of $X^{0}$ in the network. 
This Null Hypothesis is defined as a specific set of constraints on the groups that 
obey this supposedly ``normal'' behavior.
More specifically, a relevant constraint 
will be that a given feature $Z$ takes the value $Z^0$ for each of these 
groups.
Let us note $f$ the number of features that are constrained
by the Null Hypothesis.
\item Second, we create a bootstrap set of $N_B$ constrained groups by
  sampling with replacement from the data groups of nodes satisfying the constraints of
  the Null Hypothesis; we use $X$ as a generic notation for the
  bootstrap samples. In some cases, in order to obtain enough different samples,
we will need to relax some constraints of the Null Hypothesis:
such a relaxed constraint on a feature $Z$ is then written as
$Z^{0}(1-\delta)\leq Z \leq Z^{0}(1+\delta)$. The value of $\delta >0$ tunes the strength of the constraint
(the choice of $\delta$ is discussed in Section \ref{subs:tradeoff}).
The sampling procedure, based on simulated annealing, is described in details in Appendix~\ref{sec:sampling}.
\item \CHANGED{For large enough $N_B$, we estimate the distribution of each feature 
for the groups in the bootstrap set. These distributions 
describe in a fully data-driven way the ``normal behavior'' of groups in the empirical network 
under the chosen
Null Hypothesis (i.e., under this particular set of $f$ constraints).}
\item \ADDED{We select a significance level $\alpha$ for testing the Null Hypothesis,
i.e., the probability to reject the Null Hypothesis even if it is true has to be less than or equal 
to $\alpha$. $\alpha$ will also be called false alarm rate in the following. In the litterature, it is 
also called probability of false detection \cite{barnes_corrigendum:_2009}. 
Because we are dealing with observable features $Z$'s that are possibly dependent, the Bonferroni 
correction is employed: 
a significance level $\alpha' = \alpha/(F-f)$ is defined
and used to test the $F-f$ individual features that are not constrained by the Null Hypothesis
(See Section~\ref{ssec:2C}).}
\item \CHANGED{To decide whether or not we can reject the Null Hypothesis with a significance level 
$\alpha$, and
how far from the Null Hypothesis the group of interest is, a suitable divergence $d$
(defined in Section~\ref{ssec:2C}) is computed from the $Z^0$'s
and the empirical distributions of the $Z$'s for the bootstraps.
When $d=0$, the Null Hypothesis cannot be rejected with a significance level $\alpha$;
when $d$ is higher, it evaluates to what extent $X^0$ deviates
from the bootstrap samples and from the formulated Null Hypothesis,
hence from the supposed ``normal behavior''.}
\item \ADDED{As a final output, two indicators of the size of the bootstrap space are computed
(defined in \ref{ssec:outputs}) to check whether the constraints with the relaxation factor $\delta$ 
remain relevant enough for the Null Hypothesis to be tested, as further discussed in \ref{subs:tradeoff}. }
\end{enumerate}

\subsection{Normalization of features, test of individual features and choice of the divergence $d$} 
\label{ssec:2C}

Each observable $Z$ is normalized into a dimensionless quantity $z$ known as the ``Z-score'': $z=(Z-\bar{Z^*})/\sigma_{Z}^*$ where $\bar{Z^*}$ is the expected value and $\sigma_{Z}^*$ the standard
deviation of the observable $Z$ in a random graph with the same weight sequence as the empirical data. 
To estimate  these values, the following procedure is considered. Random graphs are obtained by randomly re-allocating the weights 
from the full weight sequence (including the zero weights, corresponding to absent links),
within the ensemble of possible links (i.e., pairs of nodes). 
This randomizes the degree of the nodes as well as their strengths (the strength is defined as the sum of the weights of the links of a node) 
and the local topological structures, and only preserves the weight sequence.
$\bar{Z^*}$ and $\sigma_{Z}^*$ are computed as the average and the standard deviation over the ensemble of such random graphs.
This normalization may seem arbitrary, but this mode of representation is chosen for its clarity 
(we can plot the results for all $F=7$ observables on the same scale) and, more importantly, it allows us to compare 
the results between groups of different sizes.

For each normalized observable $z$, the empirical distribution function $\hat{D}_z^b$ is derived from the bootstrap set.
As mentioned above, a statistical test is then performed 
on $X^0$ for the $z$'s to decide if $X^0$ appears as statistically far from the bootstrap set of not. 
The significance level $\alpha$ of the test cannot be used directly in this case, as we are in a situation of multiple tests
(the number of tests is $F-f$ as $f$ features are constrained by the Null Hypothesis). 
As the tests against the various features are not necessarily independent, the Bonferonni correction is used: this correction states that if
we test each feature with a significance level $\alpha' = \alpha/(F-f)$, the whole family of tests (i.e., the combined test for all the
features) holds under a significance level $\alpha$ (which would be a pessimistic, higher bound
of the true false alarm rate).
Hence, the Null Hypothesis is rejected for a specific feature $z$ if $z^0$ (the actual measured value for $X^0$)  
is outside the $1-\alpha'$ two-sided acceptance interval
for $\hat{D}_z^b$. 

We finally define a divergence $d$ quantifying if $X^0$ is far from the bootstrap set of not. 
For each observable feature $Z$, we define $d_{z}$ as the minimum distance between $z^{0}$
and the $1-\alpha'$ two-sided acceptance interval for $\hat{D}_z^b$.
If $z^0$ is in the interval, $d_Z=0$. As this interval has the meaning of an acceptance interval for the Null Hypothesis for this feature, 
$d_{z}$ measures the deviation of the observed value $Z^0$ for $X^0$ from the Null Hypothesis.

We then consider the sum $d$ of the divergences $d_{z}$ 
as the global divergence measuring to what extent we have to reject the Null Hypothesis for $X^0$:
if $d$ is larger than 0, the Null Hypothesis is rejected for the group $X^0$ with a significance level $\alpha$ and
the larger is $d$, the further away is the group $X^0$ from the Null Hypothesis. 
If $d$ equals 0 on the other hand, there is no reason to reject the Null Hypothesis under the significance 
level $\alpha$.

 \subsection{Outputs of the constrained bootstrap method}
 \label{ssec:outputs}

In classical \textit{unconstrained} bootstrap, the relevance of the test relies on an unbiased randomness in the drawing 
of the samples \cite{zoubir2004bootstrap}. 
In the present case, by imposing constraints on the bootstrap samples, some randomness is lost and this introduces possible dependencies: 
while the divergence $d$ is sufficient to summarize an unconstrained test's outcome, we need here 
to track the bias introduced by the constraints. In the following, we propose a practical way to check the validity of the procedure.

We consider two indicators to monitor the bias introduced by the constraints. 
The first one is the standard deviation $\sigma_u$ of the distribution of the number of times each node is chosen in a
bootstrap sample. It measures how uniformly a node is chosen in a bootstrap sample: the smaller is $\sigma_u$, the more the 
choice of the nodes for the bootstrap set is uniform.
The second indicator measures if nodes in $X^0$ are chosen more --\;or less\;-- often in 
the bootstrap samples than they would if there were no constraints. 
To this aim, we compare the empirical distribution of the number of nodes from $X^0$ that are in a
bootstrap sample to the theoretical distribution that would emerge if there were no constraints. This theoretical
probability distribution is the one of drawing $k$ nodes from $X^0$ after $M=|X^0|$ draws without replacement in a  
set of $V = |\mathcal{V}|$ nodes: it is given by the hypergeometric 
law $P(k)={{{M \choose k} {{V-M} \choose {M-k}}}\over {V \choose M}}$.
We thus compute the $\chi^2$ distance between the empirical distribution and the theoretical hypergeometric distribution. In order to compare 
different $\chi^2$ obtained from different bootstrap tests, each $\chi^2$ value is computed with 10 bins 
that contain at least five realisations.
An important point is that we do not use $\chi^2$ for a goodness-of-fit test. We indeed expect $\chi^2$ to increase as soon as 
we impose strong constraints on the bootstrap samples. 
Rather, we use $\chi^2$ and $\sigma_u$ as two control parameters of the ``uniform character'' of the bootstrapping procedure, 
and check that they stay reasonably small.

Overall, the final output of the proposed test is a triplet $(d,\chi^2,\sigma_u)$ that sums up the outcome of the test for $X^0$,
under the two parameters given by the significance level $\alpha$ and the relaxation factor $\delta$ for the constraints.
The larger is $d$, the further away the group is from the Null Hypothesis.
The smaller $\chi^2$ and $\sigma_u$, the less biased is the choice of the bootstrap set.

\subsection{Trade-off between the constraint(s) strength and the statistical power of the test}
\label{subs:tradeoff}

The parameter $\delta$ tunes the ``strength'' of a given constraint: the lower $\delta$, 
the stronger the constraint. Consider a very strong constraint (with a small parameter 
$\delta$). In this case, the space of possible bootstraps may be drastically 
reduced to the point where the only possible bootstraps that verify the constraint are 
very similar to the tested group $X^0$. The test will then naturally be unable to reject 
the Null Hypothesis ($d=0$) even if $X^0$ is abnormal (i.e. the test has a low statistical power)! 
In other words, consider an abnormal group $X^0$. One can always find a constraint 
(or a set of constraints) strong enough that will classify $X^0$ as normal. There is 
therefore a minimal value of $\delta$ under which the test loses its power.

On the other hand, the point of developing a method of bootstrapping under constraint   
is to test groups with highly 
specific Null Hypotheses, and to be able to understand precisely why a group is abnormal 
or not. We therefore want $\delta$ to be as small as possible in order to have bootstraps as 
representative of the Null Hypothesis as possible.

Hence, for each given constraint, there exists a trade-off value $\delta^*$ of $\delta$ 
that maximizes both the power and the precision of the test.
The existence of a threshold value $\delta^*$  for $\delta$ transposes in the existence 
of maximum authorized values $\chi^{2*}$ and $\sigma_u^*$.

In order to carry out the procedure outlined in this Section, we thus need to estimate 
$\delta^*$, $\chi^{2*}$ and $\sigma_u^*$. A theoretical estimation remains an open question~\footnote{
The underlying reason of this issue is the size of the bootstrap space, as a direct estimation 
of it is intractable because it is too huge 
(there are $C^k_n$ groups of $k$ nodes in a set of $n$ nodes). 
Instead, we use $\chi^2$ and $\sigma_u$ as two indirect measures of the size of the 
bootstrap space: the larger they are, the smaller is the bootstrap space.}. We therefore
use a controlled graph model, for different types of constraints and 
different cardinality of groups, and estimate in each case the corresponding threshold values.
Details and results are exposed in Appendix \ref{sec:ChungLu}, and we will use in the following 
the values of $\delta^*$, $\chi^{2*}$ and $\sigma_u^*$  obtained in this way.

To sum up the discussion, the test has three possible outputs:
\begin{enumerate}
 \item $d>0, \forall (\chi^{2},\sigma_u)$. 
 In this case where $d>0$, there is no need to discuss the values of $\chi^{2}$ and 
 $\sigma_u$. Indeed, even if $\chi^{2}>\chi^{2*}$ and/or $\sigma_u>\sigma_u^*$, i.e., 
 even if the bootstraps seem too similar to $X^0$, $X^0$'s behavior is still observed to be 
 different than the bootstraps: the Null Hypothesis is rejected.
 \item $d=0, \chi^{2}<\chi^{2*}, \sigma_u<\sigma_u^*$. The bootstrap space 
 is large enough, the test maintains its statistical power: the Null Hypothesis is not 
 rejected.
 \item $d=0, \chi^{2}>\chi^{2*}$ and/or $\sigma_u>\sigma_u^*$. In this case, we are 
 in the situation discussed above:  the test is not powerful enough and 
 no conclusion can be made. 
\end{enumerate}

\section{Case study: bootstrapping under constraints for specific groups of attendees in the conference}
\label{sec:5}

In order to illustrate our procedure, we consider 
a dataset describing the face-to-face proximity of individuals, collected in Salt Lake City (SLC) in November 2011 during two
co-located scientific conferences lasting five days. These conferences were jointly organised by the DPP
(Division of Plasma Physics) of the American Physical Society and the GEC (Gaseous Electronics Conference) in an attempt to bring
together both groups --\;mainly academic researchers and engineers, respectively.
A description of the context, the data collection procedure and the dataset is provided in Appendix~\ref{sec:4}. 
We provide in Table~\ref{tab:dataSLC} some basic statistics of the data. 
Note that the sum of the total number of contacts (and the total time 
of contact) within DPP and within GEC does not exactly account for the interactions for the 
conference taken as a whole (ALL), due to the interactions between DPP and GEC. 
We will consider here the aggregated
network of face-to-face proximity between individuals, in which each node represents an individual  and where
the weight of a link between two individuals gives the cumulated time they have spent in face-to-face interaction
during the conference. Moreover, we pre-process the obtained contact network by deleting links between nodes that 
correspond to an aggregated contact time of the two corresponding individuals smaller than
$1$ minute over the whole conference. 
The threshold of $1$ minute is chosen because smaller contact times can be considered as noise in the measurement, 
associated to very short contacts. We have checked that our results are robust with respect to the filtering threshold: 
similar results are obtained when thresholding at $3$ and $5$ minutes.

\begin{table}
\begin{center}
 \begin{tabular}{l|c|c|c|}
            \multicolumn{1}{c|}{} & \multicolumn{3}{c|}{SLC} \\
            \multicolumn{1}{c|}{ } & GEC & DPP & ALL \\
            \hline
            No. of tags  & 39 &  281 &320 \\\hline
            sample rate  & 12\% & 16\% & 15\% \\\hline
            No. of days &\multicolumn{3}{c|}{5} \\\hline
            No. of contacts & 1189 & 21519 & 23920 \\\hline
            Tot. time of contact (hours)& 18 & 306 & 339 \\
\end{tabular}
\end{center}
\caption{Basic statistics concerning the datasets collected in the co-located scientific conferences.}
\label{tab:dataSLC}
\end{table}

The original question of interest for the organizers of the SLC conferences is whether co-locating 
both conferences was worthwhile, i.e., whether the GEC and DPP groups mixed together.
In order to give a quantitative answer to the 
question, one needs to compare the amount of interactions between GEC and DPP to some
reference. To this aim, the proposed bootstrap method is a natural candidate.

\subsection{Choosing the groups and the Null Hypotheses}

The $F=7$ chosen 
observable features that characterize a group's ``behavior''  include features measured within the 
group ($N_{XX}$ and $T_{XX}$), features measured within the rest of the network  ($N_{RR}$ and $T_{RR}$), 
and features measuring the interaction between the group and the rest of the network 
($N_{XR}$, $T_{XR}$ and  $Q_X$). 
The terminology ``group's behavior'' is used for simplicity, but the chosen features quantify also 
the behavior of the group's complementary as well as the interaction of the group with the rest of the network.
Quantifying for instance ``GEC's behavior'' represents therefore a possible measurement 
of the mixing between both groups,
as the DPP individuals correspond precisely to the ``rest'' of the network.
The method previously exposed is a means not only to quantify, but also to validate statistically, 
the normality --\;or abnormality\;-- of GEC's behavior with respect to various Null Hypotheses. We thus 
use this method for the group of GEC individuals, taken as the specific subset of interest $X^0$ in the face-to-face
contact network between the attendees of the SLC conference. All the statistical tests reported afterwards 
are done under a significance level $\alpha = 5\%$.

\ADDED{
It is difficult to decide on only one specific Null Hypothesis that should describe the expected behavior
of a given group during the conference. There are few models for the dynamics of face-to-face contacts 
(e.g., \cite{stehle2010dynamical,Zhao2011,Starnini}), and none has been designed to 
account for all the possible features of groups in a social network, so that we can not simply compare the results
of such models with our data.
The proposed bootstrap method allows here for an interesting approach: instead of deciding on an
arbitrary Null Hypothesis, we can test the behavior of the GEC group against various Null Hypotheses
that can be formulated. The objective of the study is not merely in knowing if the GEC group was different from other
groups in the conference, but in knowing in which respect GEC is (or is not) different from other groups in a
statistically significant manner.
}

The different Null Hypotheses on which we will use the bootstrap statistical 
tests are taken as constraints on the amount of interaction involving nodes of $X$. 
We will consider several different Null Hypotheses, or sets of constraints: in each case, the Null Hypothesis
can be phrased as ``$X^0$ has a behavior compatible with a random group $X$ of nodes satisfying the 
chosen set of constraints''.
The considered sets of constraints defining these Null Hypotheses are:
\begin{itemize}
\item the size of the group is fixed, equal to the one of $X^0$;  this constraint is always active, so that there is no
effect due to the variation of the size of the group. As there are no constraints on the chosen features, $f=0$.
\item the modularity of the partition of the network between the group and its complement is equal
to the one of the partition $(X^0, {\cal V}\backslash X^0)$, up to a relaxing factor $\delta$. 
\ADDED{Note that here $f=1$ to compute $\alpha' = \alpha/(F-f)$.}
\item  constraints are put on $N_{XX}$ or $T_{XX}$, imposing
that they take the same values as respectively 
$N_{X^0X^0}$ or $T_{X^0X^0}$ 
(still in addition to the cardinality constraint).
These constraints correspond to ways of imposing a certain number 
of links or a certain amount of interaction 
within the group. \ADDED{Here again, $f=1$.}
\end{itemize}

Moreover, it is possible that all groups with a community behavior (as
given quantitatively by the seven features) could appear as
abnormal. We thus investigate the case of three other specific groups
of individuals that might \textit{a priori} present a community
behavior: the Students from DPP, i.e. attendants preparing a PhD
thesis (STP), the Juniors from DPP, i.e. researchers with less than 10
years of professional experience (JUP), and the Seniors from DPP, i.e.
researchers with more than 10 years of experience (SEP).
Table~\ref{table_obs} summarizes the measured features for the GEC,
STP, JUP and SEP.  
\CHANGED{ One could expect each of these group to
  form a community in the contact network because of the similarities
  of their members in age and professional status. When partitioning
  the network into one of these groups and its complement, the
  modularity presents indeed a high enough value.  It is therefore
  sound} to compare the tests' outputs for GEC and for these other
groups: if their behavior is similar, it could be argued that the
subgroup GEC simply behaves as if it were a subgroup of interest of
DPP, and the conclusion would be that the co-location of the
conference was an efficient way to bring together GEC and DPP.  If
instead GEC is significantly more abnormal than the three other
groups, one may doubt the efficiency of the co-location.

\begin{table}
\begin{center}
 \begin{tabular}{c|c|c|c|c|c|c|c|c}
	    Group & Cardinality & $N_{XX}$ & $N_{XR}$ & $N_{RR}$ &  $T_{XX}$ & $T_{XR}$ & $T_{RR}$ & $Q_X$\\\hline
	    GEC & 39 & 101 & 120 & 1907 & 58820 & 45740 & 947100 & 0.100\\\hline
	    STP & 106 & 384 & 850 & 894 & 252900 & 356220 & 442540 & 0.145\\\hline
	    JUP & 73 & 183 & 766 & 1179 & 97600 & 303800 & 650260 & 0.073\\\hline
	    SEP & 99 & 226 & 704 & 1198 & 124280 & 310740 & 616640 & 0.095\\
\end{tabular}
\end{center}
\vspace{-0.5cm}
\caption{The cardinality and the other seven features of the four groups under study. The temporal quantities $T_{XX}$, $T_{XR}$ and $T_{RR}$ 
are given in seconds.}
\label{table_obs}
\end{table}

Our approach is therefore to test those four groups (i.e., the group noted $X^{0}$ in the method will alternatively be 
GEC, STP, JUP, or SEP) against the same Null Hypotheses and to compare the degree with 
which the Null Hypotheses are rejected for each group.

\subsection{Results}

We first consider the simple cardinality constraint. More precisely, we test if
GEC behaves like any random group of $M=39$ individuals in the conference. As could be expected, the corresponding
Null Hypothesis is rejected, which does not come as a surprise given the quite large value of $Q_X$. In fact,
the Null Hypothesis is as well rejected for the three other groups (SEP, JUP, STP), which means that this simple
constraint does not allow us to assert if GEC behaves differently from these other groups.
Details on the procedure and its outcome are provided in Appendix~\ref{sec:card}.

\begin{figure}
\begin{center}
 \begin{minipage}{0.48\linewidth}
 \begin{center}
  \includegraphics[width=0.8\textwidth]{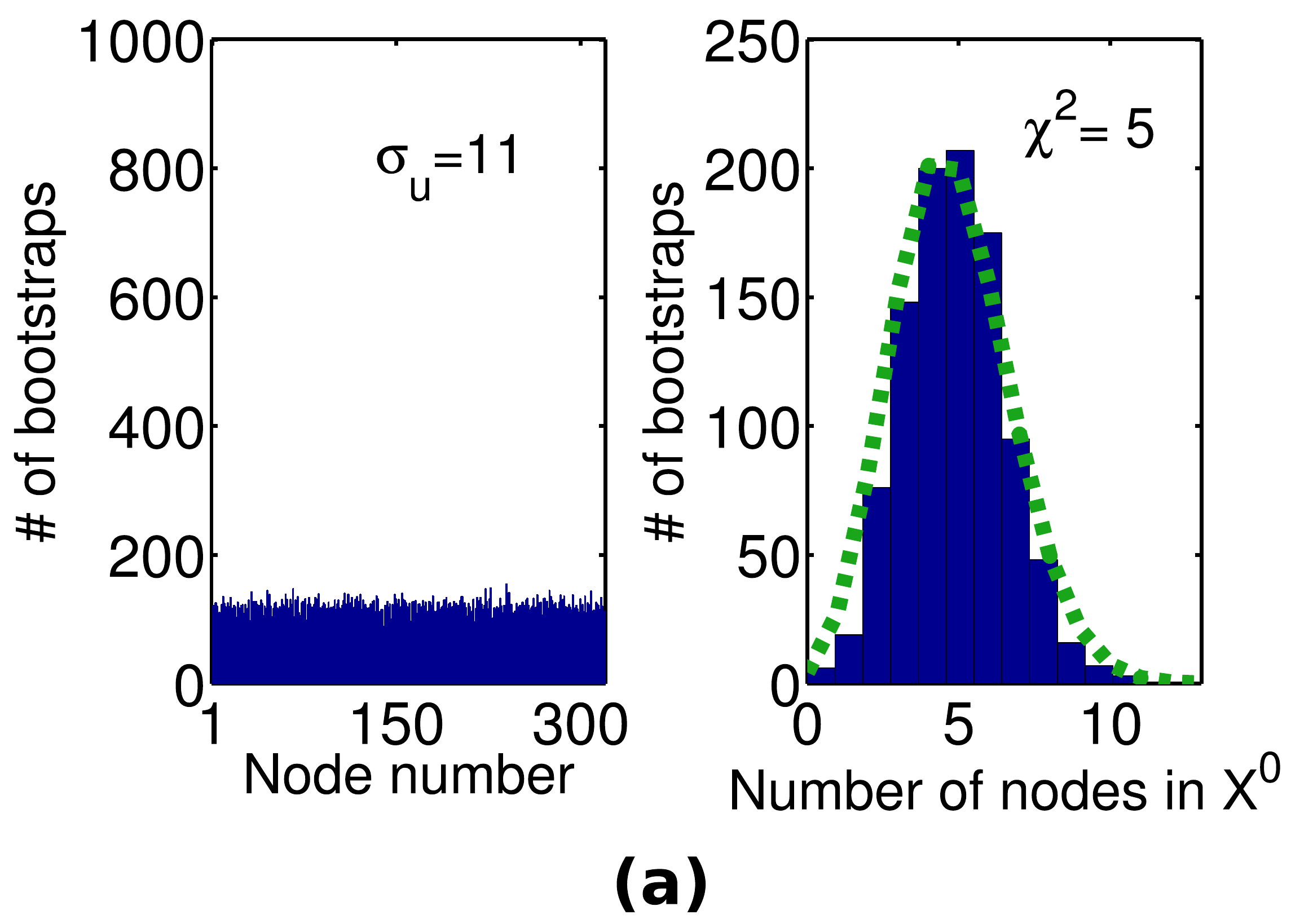}
  \end{center}
 \end{minipage}
\begin{minipage}{0.48\linewidth}
\begin{center}
\includegraphics[width=0.8\textwidth]{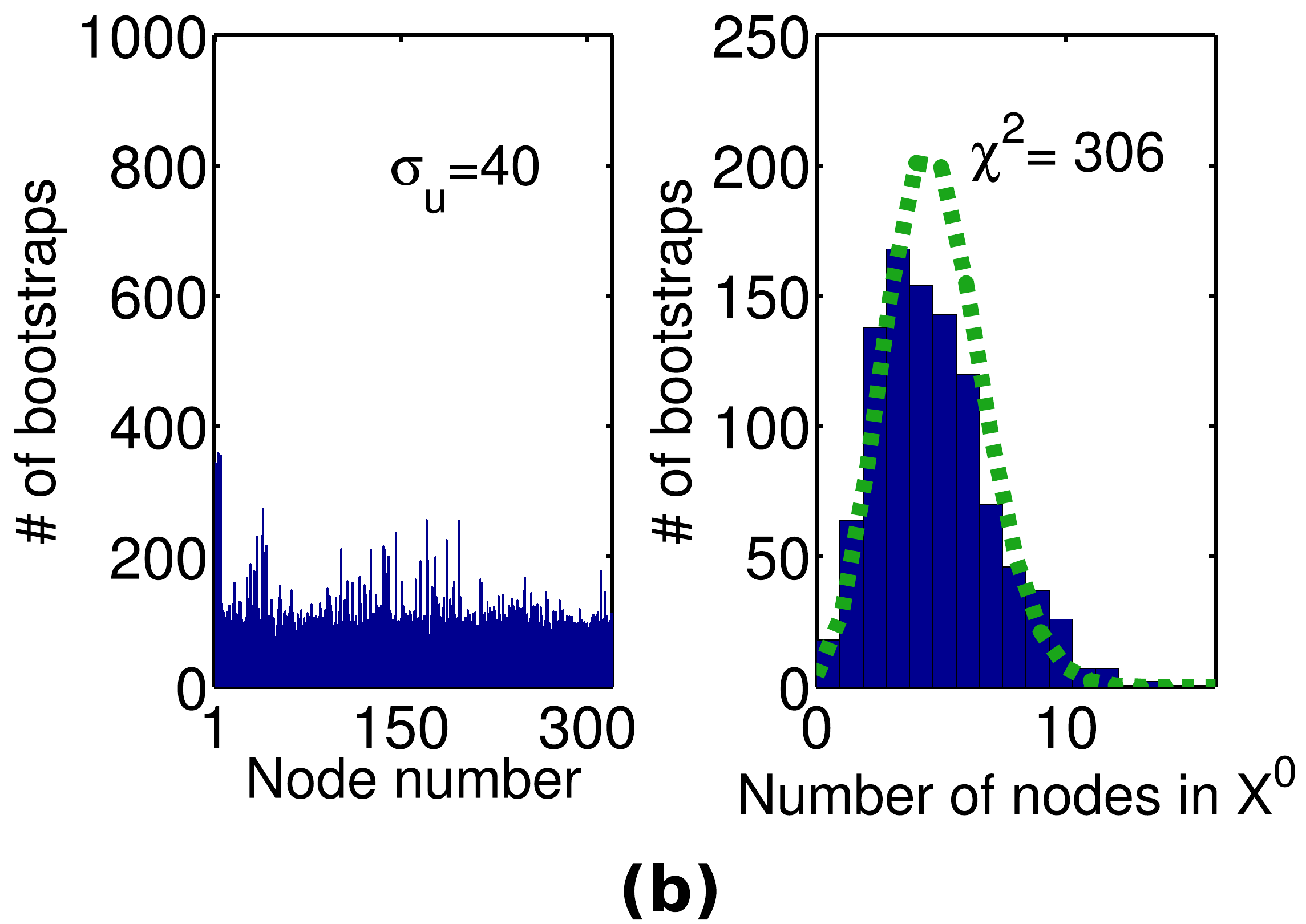}
\end{center}
 \end{minipage}
\end{center}
\caption{Outputs $\sigma_u$ and $\chi^2$ of the bootstrap method for $X^0$ = GEC for (a) test with the same cardinality
  constraint (further detailed in Appendix~\ref{sec:card}), (b) test with the constraints of same cardinality and same modularity (with $\delta=15\%$).
   Left: histogram of the number of occurrences of each
  node in the bootstrap samples and its standard deviation
  $\sigma_u$. Right: histogram of the number of $X^0$-nodes in a bootstrap
  sample with its $\chi^2$ distance from the theoretical
  hypergeometric histogram (dotted line).}
\label{histos} 
\end{figure}

\begin{figure}

\begin{center}
 \begin{minipage}{0.48\linewidth}
 \begin{center}
\includegraphics[width=0.8\textwidth]{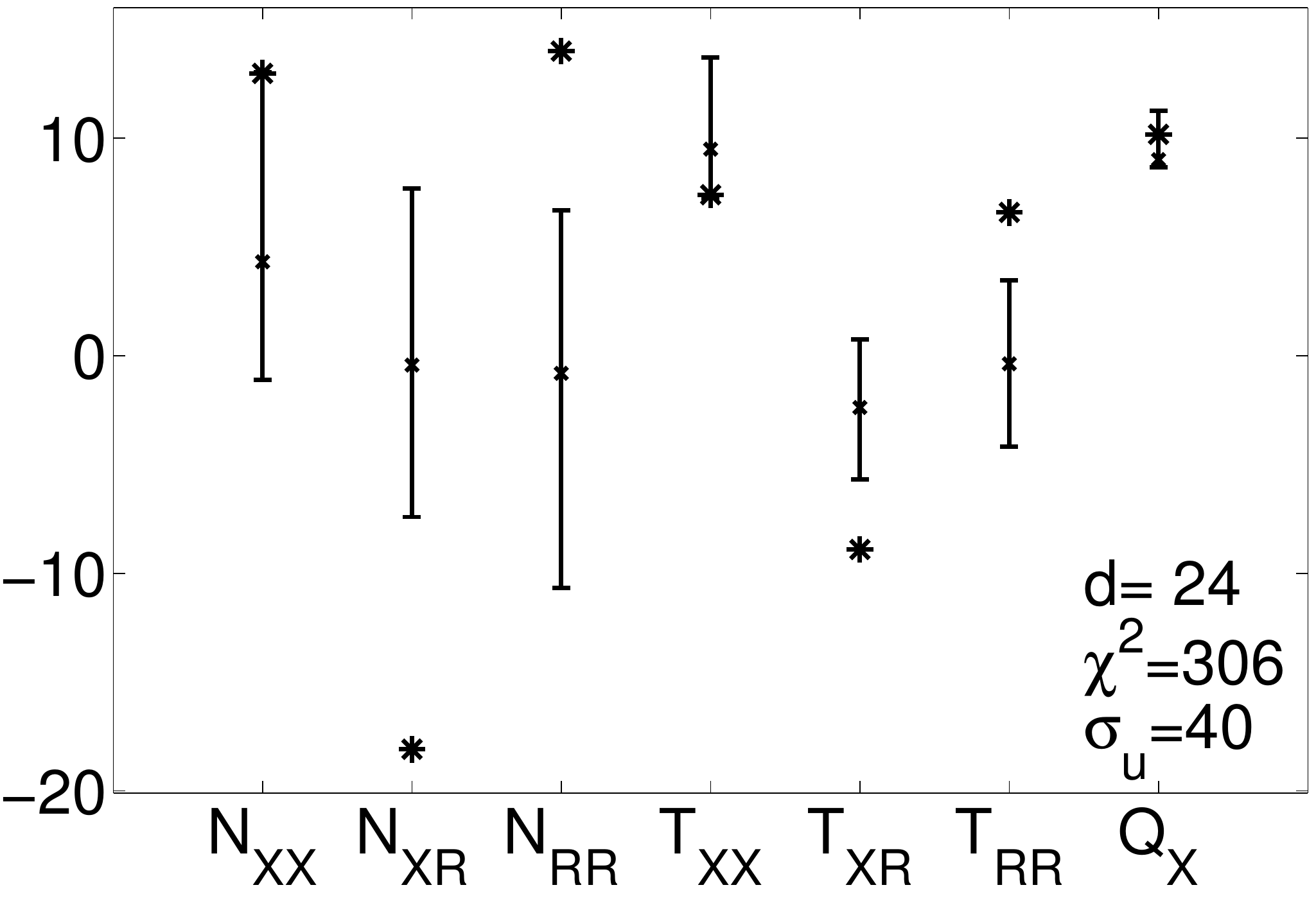} \\
 \vspace{0.5cm}
\includegraphics[width=0.8\textwidth]{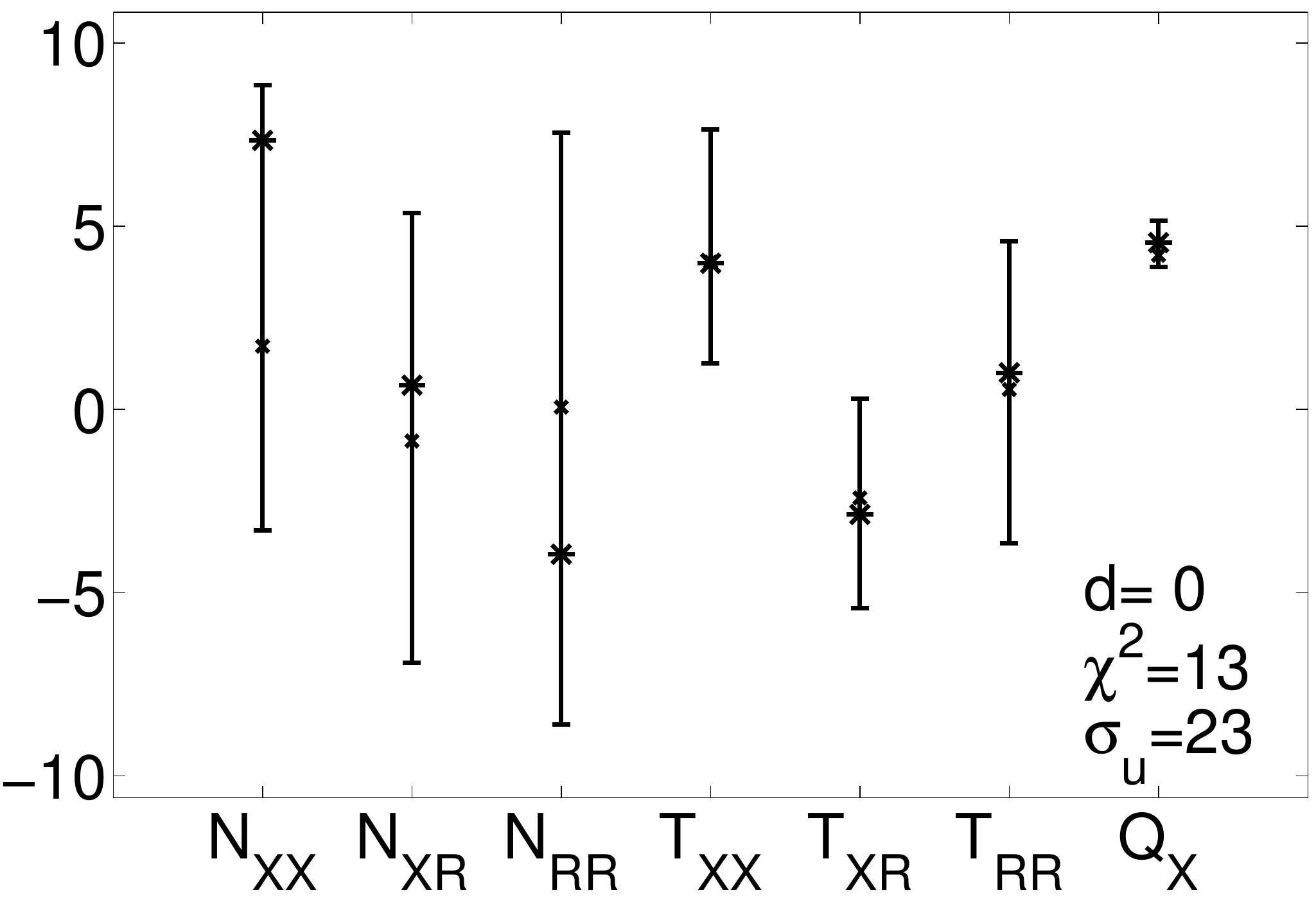}
 \end{center}
 \end{minipage}
\hfill
 \begin{minipage}{0.48\linewidth}
 \begin{center}
\includegraphics[width=0.8\textwidth]{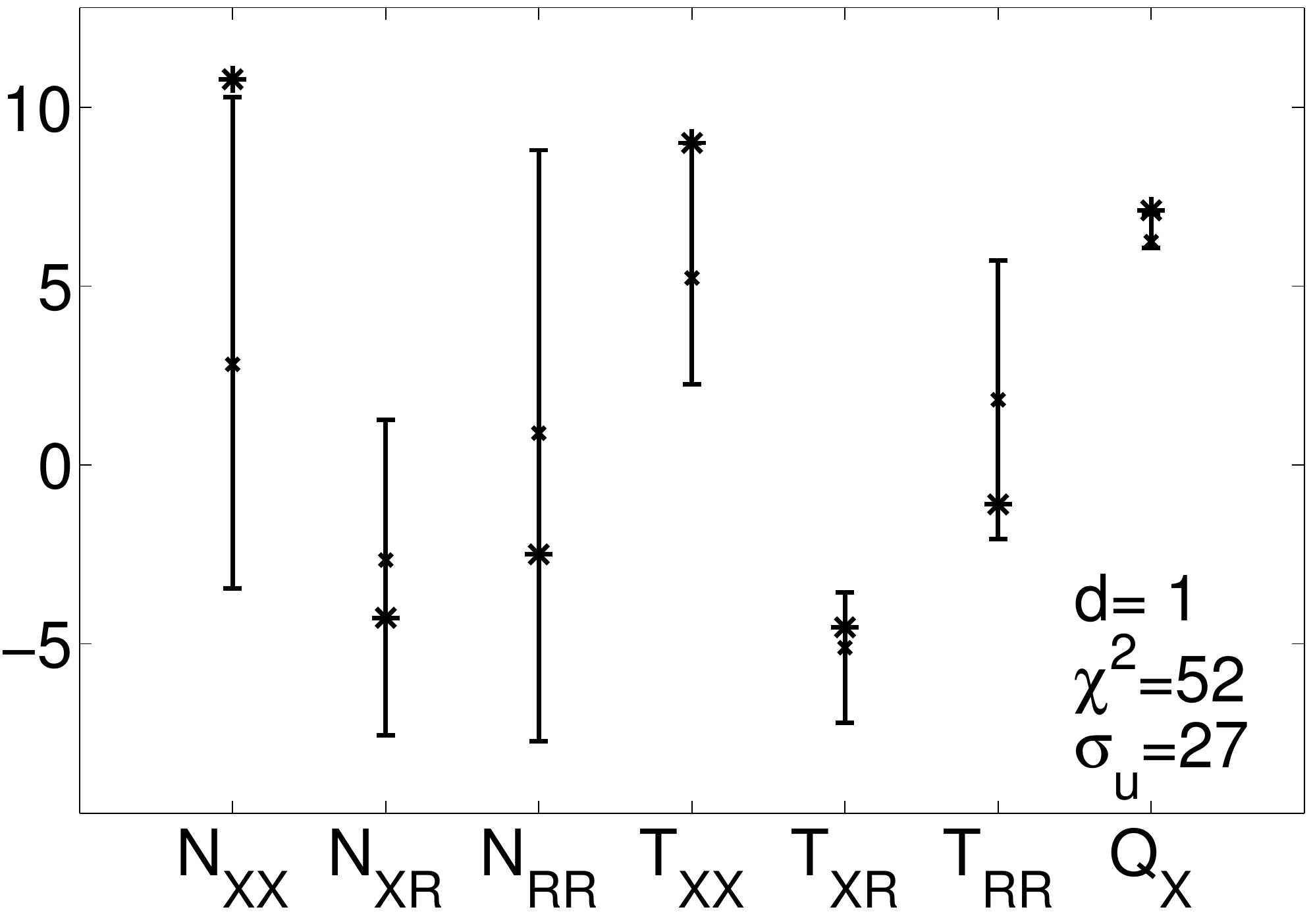}\\

 \vspace{0.5cm}
\includegraphics[width=0.8\textwidth]{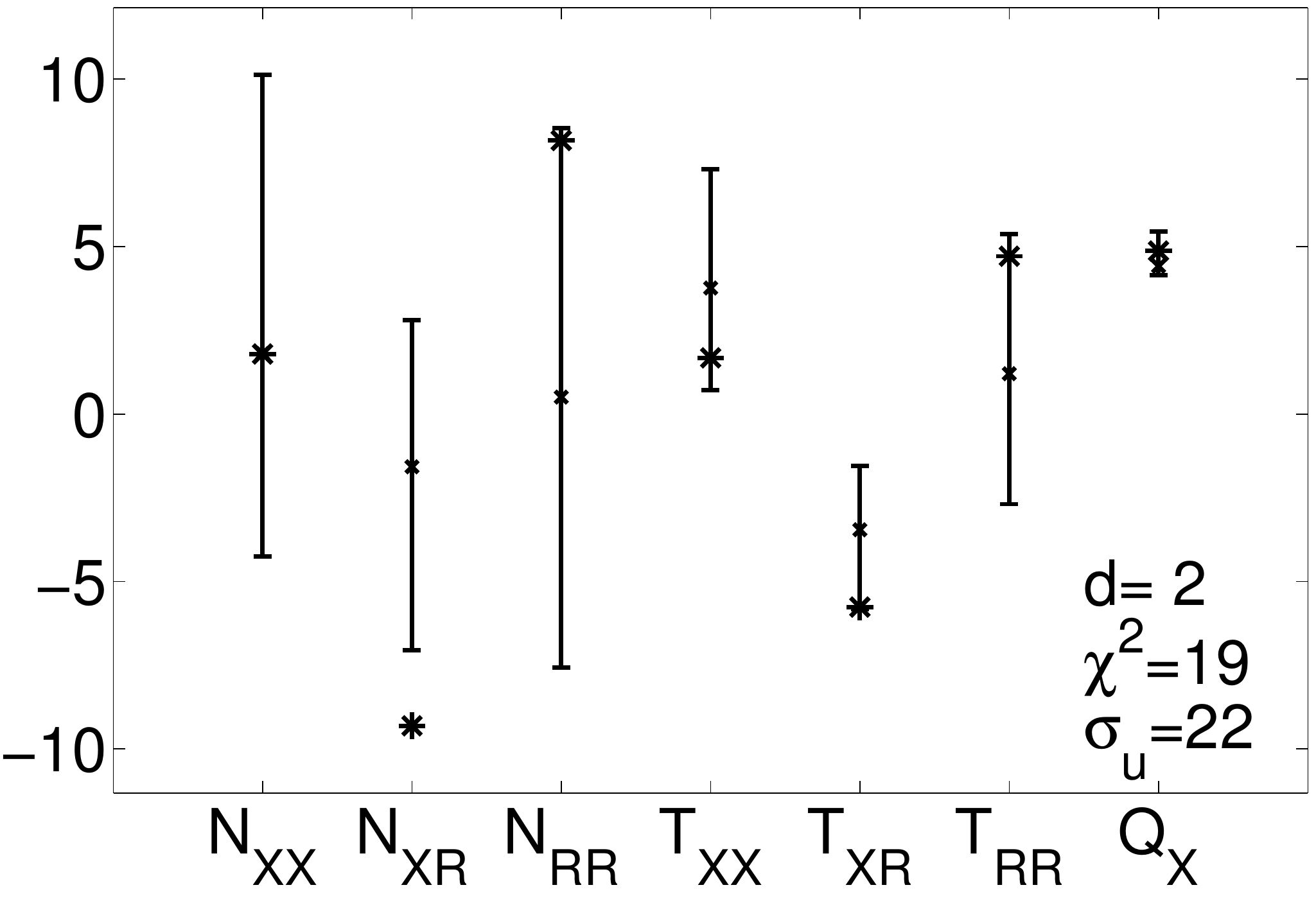}
  \end{center}
 \end{minipage}
\end{center}
\vspace{-0.5cm}
\caption{\CHANGED{Results of the test with the constraints of fixed cardinality and 
fixed modularity (with $\delta =15\%$) for the four groups: GEC (top left), STP (top right), JUP (bottom left), and SEP (bottom right).
For each $z$, the two-sided acceptance interval with $1-\alpha'$ 
significance level is in black 
and the value $z^0$ is shown as a black star. Here, $\alpha=5\%$, i.e. 
 $\alpha'=\frac{\alpha}{F-f}=\frac{0.05}{6}=0.8\%$.
For each group $X^0=$ GEC, STP, JUP and SEP, the scalar $d$ 
(bottom right hand corner of each figure) is the total 
divergence between the acceptance interval of the bootstrap samples and the real data. 
$\chi^2$ and $\sigma_u$ are the two control parameters of the size of the bootstrap space.}
 }
\label{Boxplot_MODU0} 
\end{figure}

\CHANGED{In order to better discriminate GEC's behavior from the behavior of 
other groups, we therefore turn to
more refined Null Hypotheses, i.e., with stronger constraints 
on the bootstrap samples. 
As discussed in Section~\ref{subs:tradeoff} and
Appendix~\ref{subs:CL_tradeoff}, the parameter $\delta$ is set to the threshold 
value $\delta^*$ 
corresponding to the type of constraint considered (see Table~\ref{tab:stars}).

\CHANGED{The first refined Null Hypothesis that we consider accounts for 
the high modularity of $X^0$:
does $X^{0}$ behave like any random group of nodes of same cardinality and 
modularity as $X^0$
(hence forming a community as strong as $X^0$)? This latter constraint on 
modularity is relaxed with $\delta^*=15\%$, according to the simulations of Appendix~\ref{subs:CL_tradeoff}.

Figure~\ref{histos} displays, in a) for the simple cardinality constraint and in b) for the present case,
the two histograms showing what the outputs $\sigma_u$ and $\chi^2$ aim at quantifying.
In each case, the histogram on the left hand side shows the number of times each node is chosen in the bootstrap set: 
the standard deviation $\sigma_u$ quantifies whether the choice is uniformly random or not.
On the right hand side, the distribution of the number of nodes of $X^0$ (GEC) chosen in each bootstrap sample is displayed:
the $\chi^2$ value measures the distance between the theoretical hypergeometric distribution and the actual one. 
Figure~\ref{histos}.b shows the two same 
histograms as Figure~\ref{histos}.a, but for the bootstrap samples under this 
new constraint (for $X^0$ = GEC). 
As expected, higher $\sigma_u$ and $\chi^2$ are obtained in Figure~\ref{histos}.b, yet not too large
and lower than the maximal values for this cardinality (see Table~\ref{tab:stars}-b) 
$\sigma_u^*=60$ and $\chi^{2*}=950$.}

The final outputs and results for the four studied groups are summarized in Figure~\ref{Boxplot_MODU0}. 
First, we see that the acceptance intervals are not centered around zero; 
they indeed need to be in accordance with a high modularity 
(typically: high $N_{XX}$, $T_{XX}$ and 
low $N_{XR}$ and $T_{XR}$). JUP's divergence is null, 
while the divergences for STP and SEP  
are more then ten times smaller than GEC's. This shows that GEC's behavior 
is peculiar with respect to the other groups considered, 
under the proposed Null Hypothesis.
}

Other Null Hypotheses, implying other constraints are considered:
imposing $N_{XX}=N_{X^0X^0}$ or $T_{XX}=T_{X^0X^0}$. 
These constraints are ways to impose the amount of interactions involving nodes of 
each group, respectively
in terms of numbers of contacts or of the cumulated duration of contacts inside 
the group. 
\CHANGED{Each constraint is implemented in a relaxed way, with the 
corresponding $\delta^*$ of Table~\ref{tab:stars}-b.
Results are summarized in Table~\ref{other_constraints} for these 
two other constraints
(combined in each case with the cardinality constraint). 
All outputs correspond to the type of output number 1 
($d>0, \forall (\chi^{2},\sigma_u)$) or number 2 
($d=0, \chi^{2}<\chi^{2*}, \sigma_u<\sigma_u^*$) 
discussed in 
Section~\ref{subs:tradeoff}. 
The result is that the divergence from 
the bootstrap samples is always much larger for GEC than for 
the other groups.}

Even though the modularity constraint is the most successful in
discriminating GEC from the three other groups, the other tests
show corroborative evidence of GEC's peculiar behavior. The
outputs of all the different tests are consistent, and they show not
only that GEC behaves in a peculiar fashion, but 
also in what ways GEC behaves differently. For instance,
under the constraint of fixed modularity, the acceptance intervals for 
GEC show that
it has particularly high $N_{XX}, N_{RR}, T_{RR}$ while having very
low $N_{XR}, T_{XR}$ and slightly low $T_{XX}$ features as compared to
random groups of nodes with the same modularity: the precise reasons 
for the rejection of the Null Hypothesis
are highlighted thanks to the proposed methodology.

\begin{table}
\begin{center}
 \begin{tabular}{c||c|c|c|c}
 Null Hypothesis & GEC  & STP & JUP & SEP\\\hline\hline
 No constraint (only cardinality)& (41, 5, 11)  & (15, 3, 15) & (3, 13, 14)& (9, 13, 15)\\\hline
$Q_X$ constraint with $\delta=15\%$& (24, 306, 40)  & (1, 52, 27) & (0, 13, 23)& (2, 19, 22)\\ \hline
$N_{XX}$ constraint with $\delta=5\%$& (69, 1960, 94)  & (15, 287, 97) & (4, 110, 68)& (21, 13, 23)\\\hline
$T_{XX}$ constraint with $\delta=5\%$&  (40, 277, 59)  & (7, 728, 121) & (0, 7, 52)& (15, 12, 30)\\
 \end{tabular}
\end{center}
 \caption{
 Summarized results for various sets of
   constraints. Each entry of the table gives the corresponding triplet $(d,
   \chi^2, \sigma_u)$. }
 \label{other_constraints}
\end{table}

\section{Conclusion}
\label{sec:6}
We have proposed in this work a generic method to compare the behavior
of specific groups of nodes within a given weighted complex network. 
The method is inherently flexible: depending on the
issue addressed in the data at hand, some observables and Null
Hypotheses will be more appropriate than others. We show via the
construction of a controlled model that our method is robust with
respect to random fluctuations of behavior and that it is able to 
detect abnormal ones with statistical significance.  We have shown on a new
dataset of time-resolved face-to-face human contacts collected during
two co-located conferences that the group formed by the participants to the
smaller conference could be considered as
abnormal in a statistically significant way. It had
fewer contact numbers and interaction durations with people from the
other conference, even when accounting for its organization as a group 
of high modularity. Another finding was that the mixing was better in
spaces that were shared by the two conferences.

More generally, the method we have proposed for bootstrapping and statistical
test in complex networks can be used in a broader context: it can be
applied to any type of data that can be modelled by graphs. Future work
includes applying this method for data collected 
at various times of the day. Another development would be to propose
Null Hypotheses that directly involve the dynamic behavior of groups and
not only their aggregated behavior over time.

\begin{acknowledgments}
We thank the SocioPatterns collaboration~\cite{sociopatterns} for
providing privileged access to the SocioPatterns sensing platform that
was used in collecting the contact data. A.B. 
is partially supported by FET project MULTIPLEX 317532. This work has been supported by the CNRS (PEPS ``ARDyC", 2011)
and the APS Division of Plasma Physics.
\end{acknowledgments}


\bibliography{biblio.bib}

\begin{thebibliography}{27}%
\makeatletter
\providecommand \@ifxundefined [1]{%
 \@ifx{#1\undefined}
}%
\providecommand \@ifnum [1]{%
 \ifnum #1\expandafter \@firstoftwo
 \else \expandafter \@secondoftwo
 \fi
}%
\providecommand \@ifx [1]{%
 \ifx #1\expandafter \@firstoftwo
 \else \expandafter \@secondoftwo
 \fi
}%
\providecommand \natexlab [1]{#1}%
\providecommand \enquote  [1]{``#1''}%
\providecommand \bibnamefont  [1]{#1}%
\providecommand \bibfnamefont [1]{#1}%
\providecommand \citenamefont [1]{#1}%
\providecommand \href@noop [0]{\@secondoftwo}%
\providecommand \href [0]{\begingroup \@sanitize@url \@href}%
\providecommand \@href[1]{\@@startlink{#1}\@@href}%
\providecommand \@@href[1]{\endgroup#1\@@endlink}%
\providecommand \@sanitize@url [0]{\catcode `\\12\catcode `\$12\catcode
  `\&12\catcode `\#12\catcode `\^12\catcode `\_12\catcode `\%12\relax}%
\providecommand \@@startlink[1]{}%
\providecommand \@@endlink[0]{}%
\providecommand \url  [0]{\begingroup\@sanitize@url \@url }%
\providecommand \@url [1]{\endgroup\@href {#1}{\urlprefix }}%
\providecommand \urlprefix  [0]{URL }%
\providecommand \Eprint [0]{\href }%
\providecommand \doibase [0]{http://dx.doi.org/}%
\providecommand \selectlanguage [0]{\@gobble}%
\providecommand \bibinfo  [0]{\@secondoftwo}%
\providecommand \bibfield  [0]{\@secondoftwo}%
\providecommand \translation [1]{[#1]}%
\providecommand \BibitemOpen [0]{}%
\providecommand \bibitemStop [0]{}%
\providecommand \bibitemNoStop [0]{.\EOS\space}%
\providecommand \EOS [0]{\spacefactor3000\relax}%
\providecommand \BibitemShut  [1]{\csname bibitem#1\endcsname}%
\let\auto@bib@innerbib\@empty
\bibitem [{\citenamefont {Cattuto}\ \emph {et~al.}(2010)\citenamefont
  {Cattuto}, \citenamefont {Van~den Broeck}, \citenamefont {Barrat},
  \citenamefont {Colizza}, \citenamefont {Pinton},\ and\ \citenamefont
  {Vespignani}}]{cattuto2010dynamics}%
  \BibitemOpen
  \bibfield  {author} {\bibinfo {author} {\bibfnamefont {C.}~\bibnamefont
  {Cattuto}}, \bibinfo {author} {\bibfnamefont {W.}~\bibnamefont {Van~den
  Broeck}}, \bibinfo {author} {\bibfnamefont {A.}~\bibnamefont {Barrat}},
  \bibinfo {author} {\bibfnamefont {V.}~\bibnamefont {Colizza}}, \bibinfo
  {author} {\bibfnamefont {J.}~\bibnamefont {Pinton}}, \ and\ \bibinfo {author}
  {\bibfnamefont {A.}~\bibnamefont {Vespignani}},\ }\href@noop {} {\bibfield
  {journal} {\bibinfo  {journal} {PloS one}\ }\textbf {\bibinfo {volume} {5}},\
  \bibinfo {pages} {e11596} (\bibinfo {year} {2010})}\BibitemShut {NoStop}%
\bibitem [{\citenamefont {Eagle}\ and\ \citenamefont
  {Pentland}(2006)}]{eagle2006reality}%
  \BibitemOpen
  \bibfield  {author} {\bibinfo {author} {\bibfnamefont {N.}~\bibnamefont
  {Eagle}}\ and\ \bibinfo {author} {\bibfnamefont {A.}~\bibnamefont
  {Pentland}},\ }\href@noop {} {\bibfield  {journal} {\bibinfo  {journal}
  {Personal and Ubiquitous Computing}\ }\textbf {\bibinfo {volume} {10}},\
  \bibinfo {pages} {255} (\bibinfo {year} {2006})}\BibitemShut {NoStop}%
\bibitem [{\citenamefont {Hui}\ \emph {et~al.}(2005)\citenamefont {Hui},
  \citenamefont {Chaintreau}, \citenamefont {Scott}, \citenamefont {Gass},
  \citenamefont {Crowcroft},\ and\ \citenamefont {Diot}}]{hui2005pocket}%
  \BibitemOpen
  \bibfield  {author} {\bibinfo {author} {\bibfnamefont {P.}~\bibnamefont
  {Hui}}, \bibinfo {author} {\bibfnamefont {A.}~\bibnamefont {Chaintreau}},
  \bibinfo {author} {\bibfnamefont {J.}~\bibnamefont {Scott}}, \bibinfo
  {author} {\bibfnamefont {R.}~\bibnamefont {Gass}}, \bibinfo {author}
  {\bibfnamefont {J.}~\bibnamefont {Crowcroft}}, \ and\ \bibinfo {author}
  {\bibfnamefont {C.}~\bibnamefont {Diot}},\ }in\ \href@noop {} {\emph
  {\bibinfo {booktitle} {Proceedings of the 2005 ACM SIGCOMM workshop on
  Delay-tolerant networking}}}\ (\bibinfo {organization} {ACM},\ \bibinfo
  {year} {2005})\ pp.\ \bibinfo {pages} {244--251}\BibitemShut {NoStop}%
\bibitem [{\citenamefont {Salath{\'e}}\ \emph {et~al.}(2010)\citenamefont
  {Salath{\'e}}, \citenamefont {Kazandjieva}, \citenamefont {Lee},
  \citenamefont {Levis}, \citenamefont {Feldman},\ and\ \citenamefont
  {Jones}}]{salathe2010high}%
  \BibitemOpen
  \bibfield  {author} {\bibinfo {author} {\bibfnamefont {M.}~\bibnamefont
  {Salath{\'e}}}, \bibinfo {author} {\bibfnamefont {M.}~\bibnamefont
  {Kazandjieva}}, \bibinfo {author} {\bibfnamefont {J.}~\bibnamefont {Lee}},
  \bibinfo {author} {\bibfnamefont {P.}~\bibnamefont {Levis}}, \bibinfo
  {author} {\bibfnamefont {M.}~\bibnamefont {Feldman}}, \ and\ \bibinfo
  {author} {\bibfnamefont {J.}~\bibnamefont {Jones}},\ }\href@noop {}
  {\bibfield  {journal} {\bibinfo  {journal} {Proceedings of the National
  Academy of Sciences}\ }\textbf {\bibinfo {volume} {107}},\ \bibinfo {pages}
  {22020} (\bibinfo {year} {2010})}\BibitemShut {NoStop}%
\bibitem [{\citenamefont {Efron}(1982)}]{efron1982jackknife}%
  \BibitemOpen
  \bibfield  {author} {\bibinfo {author} {\bibfnamefont {B.}~\bibnamefont
  {Efron}},\ }\href@noop {} {\emph {\bibinfo {title} {The jackknife, the
  bootstrap, and other resampling plans}}},\ Vol.~\bibinfo {volume} {38}\
  (\bibinfo  {publisher} {Society for Industrial and Applied Mathematics
  Philadelphia},\ \bibinfo {year} {1982})\BibitemShut {NoStop}%
\bibitem [{\citenamefont {Zoubir}\ and\ \citenamefont
  {Iskander}(2004)}]{zoubir2004bootstrap}%
  \BibitemOpen
  \bibfield  {author} {\bibinfo {author} {\bibfnamefont {A.}~\bibnamefont
  {Zoubir}}\ and\ \bibinfo {author} {\bibfnamefont {D.}~\bibnamefont
  {Iskander}},\ }\href@noop {} {\emph {\bibinfo {title} {Bootstrap techniques
  for signal processing}}}\ (\bibinfo  {publisher} {Cambridge University
  Press},\ \bibinfo {year} {2004})\BibitemShut {NoStop}%
\bibitem [{\citenamefont {Eldardiry}\ and\ \citenamefont
  {Neville}(2008)}]{eldardiry2008resampling}%
  \BibitemOpen
  \bibfield  {author} {\bibinfo {author} {\bibfnamefont {H.}~\bibnamefont
  {Eldardiry}}\ and\ \bibinfo {author} {\bibfnamefont {J.}~\bibnamefont
  {Neville}},\ }in\ \href@noop {} {\emph {\bibinfo {booktitle} {Proceedings of
  the 2nd SNA Workshop, 14th ACM SIGKDD Conference on Knowledge Discovery and
  Data Mining}}}\ (\bibinfo {year} {2008})\BibitemShut {NoStop}%
\bibitem [{\citenamefont {Ying}\ and\ \citenamefont
  {Wu}(2009)}]{ying2009graph}%
  \BibitemOpen
  \bibfield  {author} {\bibinfo {author} {\bibfnamefont {X.}~\bibnamefont
  {Ying}}\ and\ \bibinfo {author} {\bibfnamefont {X.}~\bibnamefont {Wu}},\ }in\
  \href@noop {} {\emph {\bibinfo {booktitle} {Proc. of the 9th SIAM Conference
  on Data Mining}}}\ (\bibinfo {year} {2009})\BibitemShut {NoStop}%
\bibitem [{\citenamefont {Drummond}\ and\ \citenamefont
  {Rambaut}(2007)}]{drummond2007beast}%
  \BibitemOpen
  \bibfield  {author} {\bibinfo {author} {\bibfnamefont {A.}~\bibnamefont
  {Drummond}}\ and\ \bibinfo {author} {\bibfnamefont {A.}~\bibnamefont
  {Rambaut}},\ }\href@noop {} {\bibfield  {journal} {\bibinfo  {journal} {BMC
  evolutionary biology}\ }\textbf {\bibinfo {volume} {7}},\ \bibinfo {pages}
  {214} (\bibinfo {year} {2007})}\BibitemShut {NoStop}%
\bibitem [{\citenamefont {Friedman}\ \emph {et~al.}(1999)\citenamefont
  {Friedman}, \citenamefont {Goldszmidt},\ and\ \citenamefont
  {Wyner}}]{friedman1999data}%
  \BibitemOpen
  \bibfield  {author} {\bibinfo {author} {\bibfnamefont {N.}~\bibnamefont
  {Friedman}}, \bibinfo {author} {\bibfnamefont {M.}~\bibnamefont
  {Goldszmidt}}, \ and\ \bibinfo {author} {\bibfnamefont {A.}~\bibnamefont
  {Wyner}},\ }in\ \href@noop {} {\emph {\bibinfo {booktitle} {Proceedings of
  the Fifteenth conference on Uncertainty in artificial intelligence}}}\
  (\bibinfo {organization} {Morgan Kaufmann Publishers Inc.},\ \bibinfo {year}
  {1999})\ pp.\ \bibinfo {pages} {196--205}\BibitemShut {NoStop}%
\bibitem [{\citenamefont {Fortunato}(2010)}]{Fortunato2010}%
  \BibitemOpen
  \bibfield  {author} {\bibinfo {author} {\bibfnamefont {S.}~\bibnamefont
  {Fortunato}},\ }\href@noop {} {\bibfield  {journal} {\bibinfo  {journal}
  {Physics Reports}\ }\textbf {\bibinfo {volume} {486}},\ \bibinfo {pages} {75}
  (\bibinfo {year} {2010})}\BibitemShut {NoStop}%
\bibitem [{\citenamefont {Lancichinetti}\ \emph {et~al.}(2010)\citenamefont
  {Lancichinetti}, \citenamefont {Radicchi},\ and\ \citenamefont
  {Ramasco}}]{Lancichinetti2010}%
  \BibitemOpen
  \bibfield  {author} {\bibinfo {author} {\bibfnamefont {A.}~\bibnamefont
  {Lancichinetti}}, \bibinfo {author} {\bibfnamefont {F.}~\bibnamefont
  {Radicchi}}, \ and\ \bibinfo {author} {\bibfnamefont {J.~J.}\ \bibnamefont
  {Ramasco}},\ }\href {\doibase 10.1103/PhysRevE.81.046110} {\bibfield
  {journal} {\bibinfo  {journal} {Phys. Rev. E}\ }\textbf {\bibinfo {volume}
  {81}},\ \bibinfo {pages} {046110} (\bibinfo {year} {2010})}\BibitemShut
  {NoStop}%
\bibitem [{\citenamefont {Rosvall}\ and\ \citenamefont
  {Bergstrom}(2010)}]{rosvall2010mapping}%
  \BibitemOpen
  \bibfield  {author} {\bibinfo {author} {\bibfnamefont {M.}~\bibnamefont
  {Rosvall}}\ and\ \bibinfo {author} {\bibfnamefont {C.}~\bibnamefont
  {Bergstrom}},\ }\href@noop {} {\bibfield  {journal} {\bibinfo  {journal}
  {PloS one}\ }\textbf {\bibinfo {volume} {5}},\ \bibinfo {pages} {e8694}
  (\bibinfo {year} {2010})}\BibitemShut {NoStop}%
\bibitem [{\citenamefont {Newman}(2004)}]{newman2004analysis}%
  \BibitemOpen
  \bibfield  {author} {\bibinfo {author} {\bibfnamefont {M.}~\bibnamefont
  {Newman}},\ }\href@noop {} {\bibfield  {journal} {\bibinfo  {journal}
  {Physical Review E}\ }\textbf {\bibinfo {volume} {70}},\ \bibinfo {pages}
  {056131} (\bibinfo {year} {2004})}\BibitemShut {NoStop}%
\bibitem [{Note1()}]{Note1}%
  \BibitemOpen
  \bibinfo {note} {{The modularity is lower than $0.5$ when there are two
  groups because the modularity of a partition in $K$ groups is known to be
  bounded by $1-1/K$~\cite {mieghem2011graph}.}}\BibitemShut {Stop}%
\bibitem [{\citenamefont {Barnes}\ \emph {et~al.}(2009)\citenamefont {Barnes},
  \citenamefont {Schultz}, \citenamefont {Gruntfest}, \citenamefont {Hayden},\
  and\ \citenamefont {Benight}}]{barnes_corrigendum:_2009}%
  \BibitemOpen
  \bibfield  {author} {\bibinfo {author} {\bibfnamefont {L.~R.}\ \bibnamefont
  {Barnes}}, \bibinfo {author} {\bibfnamefont {D.~M.}\ \bibnamefont {Schultz}},
  \bibinfo {author} {\bibfnamefont {E.~C.}\ \bibnamefont {Gruntfest}}, \bibinfo
  {author} {\bibfnamefont {M.~H.}\ \bibnamefont {Hayden}}, \ and\ \bibinfo
  {author} {\bibfnamefont {C.~C.}\ \bibnamefont {Benight}},\ }\href {\doibase
  10.1175/2009WAF2222300.1} {\bibfield  {journal} {\bibinfo  {journal} {Weather
  and Forecasting}\ }\textbf {\bibinfo {volume} {24}},\ \bibinfo {pages} {1452}
  (\bibinfo {year} {2009})}\BibitemShut {NoStop}%
\bibitem [{Note2()}]{Note2}%
  \BibitemOpen
  \bibinfo {note} {The underlying reason of this issue is the size of the
  bootstrap space, as a direct estimation of it is intractable because it is
  too huge (there are $C^k_n$ groups of $k$ nodes in a set of $n$ nodes).
  Instead, we use $\chi ^2$ and $\sigma _u$ as two indirect measures of the
  size of the bootstrap space: the larger they are, the smaller is the
  bootstrap space.}\BibitemShut {Stop}%
\bibitem [{\citenamefont {Stehl{\'e}}\ \emph {et~al.}(2010)\citenamefont
  {Stehl{\'e}}, \citenamefont {Barrat},\ and\ \citenamefont
  {Bianconi}}]{stehle2010dynamical}%
  \BibitemOpen
  \bibfield  {author} {\bibinfo {author} {\bibfnamefont {J.}~\bibnamefont
  {Stehl{\'e}}}, \bibinfo {author} {\bibfnamefont {A.}~\bibnamefont {Barrat}},
  \ and\ \bibinfo {author} {\bibfnamefont {G.}~\bibnamefont {Bianconi}},\
  }\href@noop {} {\bibfield  {journal} {\bibinfo  {journal} {Physical review
  E}\ }\textbf {\bibinfo {volume} {81}},\ \bibinfo {pages} {035101} (\bibinfo
  {year} {2010})}\BibitemShut {NoStop}%
\bibitem [{\citenamefont {Zhao}\ \emph {et~al.}(2011)\citenamefont {Zhao},
  \citenamefont {Stehl\'e}, \citenamefont {Bianconi},\ and\ \citenamefont
  {Barrat}}]{Zhao2011}%
  \BibitemOpen
  \bibfield  {author} {\bibinfo {author} {\bibfnamefont {K.}~\bibnamefont
  {Zhao}}, \bibinfo {author} {\bibfnamefont {J.}~\bibnamefont {Stehl\'e}},
  \bibinfo {author} {\bibfnamefont {G.}~\bibnamefont {Bianconi}}, \ and\
  \bibinfo {author} {\bibfnamefont {A.}~\bibnamefont {Barrat}},\ }\href
  {\doibase 10.1103/PhysRevE.83.056109} {\bibfield  {journal} {\bibinfo
  {journal} {Phys. Rev. E}\ }\textbf {\bibinfo {volume} {83}},\ \bibinfo
  {pages} {056109} (\bibinfo {year} {2011})}\BibitemShut {NoStop}%
\bibitem [{\citenamefont {Starnini}\ \emph {et~al.}(2013)\citenamefont
  {Starnini}, \citenamefont {Baronchelli},\ and\ \citenamefont
  {Pastor-Satorras}}]{Starnini}%
  \BibitemOpen
  \bibfield  {author} {\bibinfo {author} {\bibfnamefont {M.}~\bibnamefont
  {Starnini}}, \bibinfo {author} {\bibfnamefont {A.}~\bibnamefont
  {Baronchelli}}, \ and\ \bibinfo {author} {\bibfnamefont {R.}~\bibnamefont
  {Pastor-Satorras}},\ }\href {\doibase 10.1103/PhysRevLett.110.168701}
  {\bibfield  {journal} {\bibinfo  {journal} {Phys. Rev. Lett.}\ }\textbf
  {\bibinfo {volume} {110}},\ \bibinfo {pages} {168701} (\bibinfo {year}
  {2013})}\BibitemShut {NoStop}%
\bibitem [{\citenamefont {www.sociopatterns.org}()}]{sociopatterns}%
  \BibitemOpen
  \bibfield  {author} {\bibinfo {author} {\bibnamefont
  {www.sociopatterns.org}},\ }\href {www.sociopatterns.org} {\ }\BibitemShut
  {NoStop}%
\bibitem [{\citenamefont {Van~Mieghem}(2011)}]{mieghem2011graph}%
  \BibitemOpen
  \bibfield  {author} {\bibinfo {author} {\bibfnamefont {P.}~\bibnamefont
  {Van~Mieghem}},\ }\href@noop {} {\emph {\bibinfo {title} {Graph spectra for
  complex networks}}}\ (\bibinfo  {publisher} {Cambridge University Press},\
  \bibinfo {year} {2011})\BibitemShut {NoStop}%
\bibitem [{\citenamefont {Brooks}\ and\ \citenamefont {Morgan}(1995)}]{1995}%
  \BibitemOpen
  \bibfield  {author} {\bibinfo {author} {\bibfnamefont {S.~P.}\ \bibnamefont
  {Brooks}}\ and\ \bibinfo {author} {\bibfnamefont {B.~J.~T.}\ \bibnamefont
  {Morgan}},\ }\href {http://www.jstor.org/stable/2348448} {\bibfield
  {journal} {\bibinfo  {journal} {Journal of the Royal Statistical Society.
  Series D (The Statistician)}\ }\textbf {\bibinfo {volume} {44}},\ \bibinfo
  {pages} {pp. 241} (\bibinfo {year} {1995})}\BibitemShut {NoStop}%
\bibitem [{\citenamefont {Chung}\ and\ \citenamefont
  {Lu}(2002)}]{chung2002average}%
  \BibitemOpen
  \bibfield  {author} {\bibinfo {author} {\bibfnamefont {F.}~\bibnamefont
  {Chung}}\ and\ \bibinfo {author} {\bibfnamefont {L.}~\bibnamefont {Lu}},\
  }\href@noop {} {\bibfield  {journal} {\bibinfo  {journal} {Proceedings of the
  National Academy of Sciences}\ }\textbf {\bibinfo {volume} {99}},\ \bibinfo
  {pages} {15879} (\bibinfo {year} {2002})}\BibitemShut {NoStop}%
\bibitem [{\citenamefont {Miller}\ and\ \citenamefont
  {Hagberg}(2011)}]{miller2011efficient}%
  \BibitemOpen
  \bibfield  {author} {\bibinfo {author} {\bibfnamefont {J.}~\bibnamefont
  {Miller}}\ and\ \bibinfo {author} {\bibfnamefont {A.}~\bibnamefont
  {Hagberg}},\ }\href@noop {} {\bibfield  {journal} {\bibinfo  {journal}
  {Algorithms and Models for the Web Graph}\ ,\ \bibinfo {pages} {115}}
  (\bibinfo {year} {2011})}\BibitemShut {NoStop}%
\bibitem [{\citenamefont {Barrat}\ \emph {et~al.}(2004)\citenamefont {Barrat},
  \citenamefont {Barth\'elemy}, \citenamefont {Pastor-Satorras},\ and\
  \citenamefont {Vespignani}}]{Barrat2004}%
  \BibitemOpen
  \bibfield  {author} {\bibinfo {author} {\bibfnamefont {A.}~\bibnamefont
  {Barrat}}, \bibinfo {author} {\bibfnamefont {M.}~\bibnamefont
  {Barth\'elemy}}, \bibinfo {author} {\bibfnamefont {R.}~\bibnamefont
  {Pastor-Satorras}}, \ and\ \bibinfo {author} {\bibfnamefont {A.}~\bibnamefont
  {Vespignani}},\ }\href@noop {} {\bibfield  {journal} {\bibinfo  {journal}
  {Proc. Natl. Acad. Sci. (USA)}\ }\textbf {\bibinfo {volume} {101}},\ \bibinfo
  {pages} {3747} (\bibinfo {year} {2004})}\BibitemShut {NoStop}%
\bibitem [{\citenamefont {Isella}\ \emph {et~al.}(2011)\citenamefont {Isella},
  \citenamefont {Stehl{\'e}}, \citenamefont {Barrat}, \citenamefont {Cattuto},
  \citenamefont {Pinton},\ and\ \citenamefont {Van~den Broeck}}]{isella2011s}%
  \BibitemOpen
  \bibfield  {author} {\bibinfo {author} {\bibfnamefont {L.}~\bibnamefont
  {Isella}}, \bibinfo {author} {\bibfnamefont {J.}~\bibnamefont {Stehl{\'e}}},
  \bibinfo {author} {\bibfnamefont {A.}~\bibnamefont {Barrat}}, \bibinfo
  {author} {\bibfnamefont {C.}~\bibnamefont {Cattuto}}, \bibinfo {author}
  {\bibfnamefont {J.}~\bibnamefont {Pinton}}, \ and\ \bibinfo {author}
  {\bibfnamefont {W.}~\bibnamefont {Van~den Broeck}},\ }\href@noop {}
  {\bibfield  {journal} {\bibinfo  {journal} {Journal of theoretical biology}\
  }\textbf {\bibinfo {volume} {271}},\ \bibinfo {pages} {166} (\bibinfo {year}
  {2011})}\BibitemShut {NoStop}%
\end{thebibliography}%

\newpage

\appendix

\section{Sampling method to create constrained bootstraps}
\label{sec:sampling}

A key technical point is that the sampling method should allow
us to draw sets of nodes that satisfy the chosen constraints.
The simplest of the constraints is the cardinality constraint \CHANGED{that sets the size of the group under study.
In this case, the cardinality of all bootstrapped groups is simply set to match $X^0$'s} so that there
is no discrepancy in the features because of different sizes of the groups $X$.
This constraint is trivially achieved: for each bootstrap sample, we randomly draw
nodes from the network (without replacement) until the size of the bootstrapped group
reaches $X^0$'s size.

Other Null Hypotheses lead us to impose stronger constraints
by requiring $f$ observables to be the same in $X$ as in $X^0$.
For example, a possible constraint is to have the ``same $N_{XX}$'', in which we
impose in addition that each bootstrap sample has the
same number of internal links than $X^0$.
As mentioned in the main text, each constraint on a feature $Z$ can be implemented sharply or with a relaxation
factor $\delta$, so that the constrained feature satisfies $Z^{0}(1-\delta)\leq Z \leq Z^{0}(1+\delta)$.

Technically, a simulated annealing algorithm~\cite{1995} is employed as follows in
order to draw each bootstrap sample so that it satisfies the constraints.
Let us start with a random set of nodes $X$, with the same cardinality as $X^0$.
The cost $C$ of $X$ is defined as the absolute difference between the value of $Z$ in the current group and $Z^0$.
An auxiliary ``temperature'' $T$ is set to start at a given value \ADDED{(here T=0.5)}.
At each step of the simulated annealing procedure, we
keep some of the nodes of the current group $X$ and change the rest
\ADDED{(more precisely, we attempt to change $\mbox{min}(M\times|r|\times T,M)$ nodes out of the $M$ nodes of the group,
where $r$ is a normally distributed random variable of mean $0$ and variance $1$)}.
If the cost $C'$ of the new group is lower than $C$, we accept the change. If instead $C' > C$, we accept the change with
probability
$p=\mbox{min} \left(\exp\left(\frac{C-C'}{T}\right),1 \right)$.
When the cost does not decrease during several attempts, we lower the auxiliary
temperature \ADDED{($T\leftarrow0.85T$)} and start the whole process again.
We stop the algorithm as soon as $X$ satisfies the constraint (as soon as $C=0$ for a sharp constraint, or
when $Z$ is between $Z^{0}(1-\delta)$ and $Z^{0}(1+\delta)$ for a relaxed constraint).

This process is repeated $N_B$ times to obtain the whole bootstrap set.
\\

\section{Controlled study on weighted random graphs}
\label{sec:ChungLu}

We perform here a validation of our methodology and the tuning of the parameters using controlled graphs.
We first present the procedure used to generate weighted Chung-Lu graphs in which we control the degree sequence
as well as the correlations between degrees and weights.
We then use such graphs to check whether the statistical test described in Section \ref{sec:2} 
has the expected false alarm rate $\alpha$.
Then, we empirically estimate $\delta^*$, $\chi^{2*}$ and $\sigma_u^*$ for 
different types of constraints and 
different cardinality of groups on this controlled model.

\subsection{Weighted Chung-Lu graphs}

A Chung-Lu graph~\cite{chung2002average, miller2011efficient} is a
random graph with a given expected degree sequence
$\left(k_i\right)_{i=1, ..., V}$. In such a graph, 
the probability that a given edge (connecting nodes 
$i$ and $j$) exists is given by min$(1,{k_i k_j}/{2W})$, where
$W=\frac{1}{2}\sum_i k_i$ is the expected total number of edges.

\begin{figure}
\begin{center}
\includegraphics[width=0.55\textwidth]{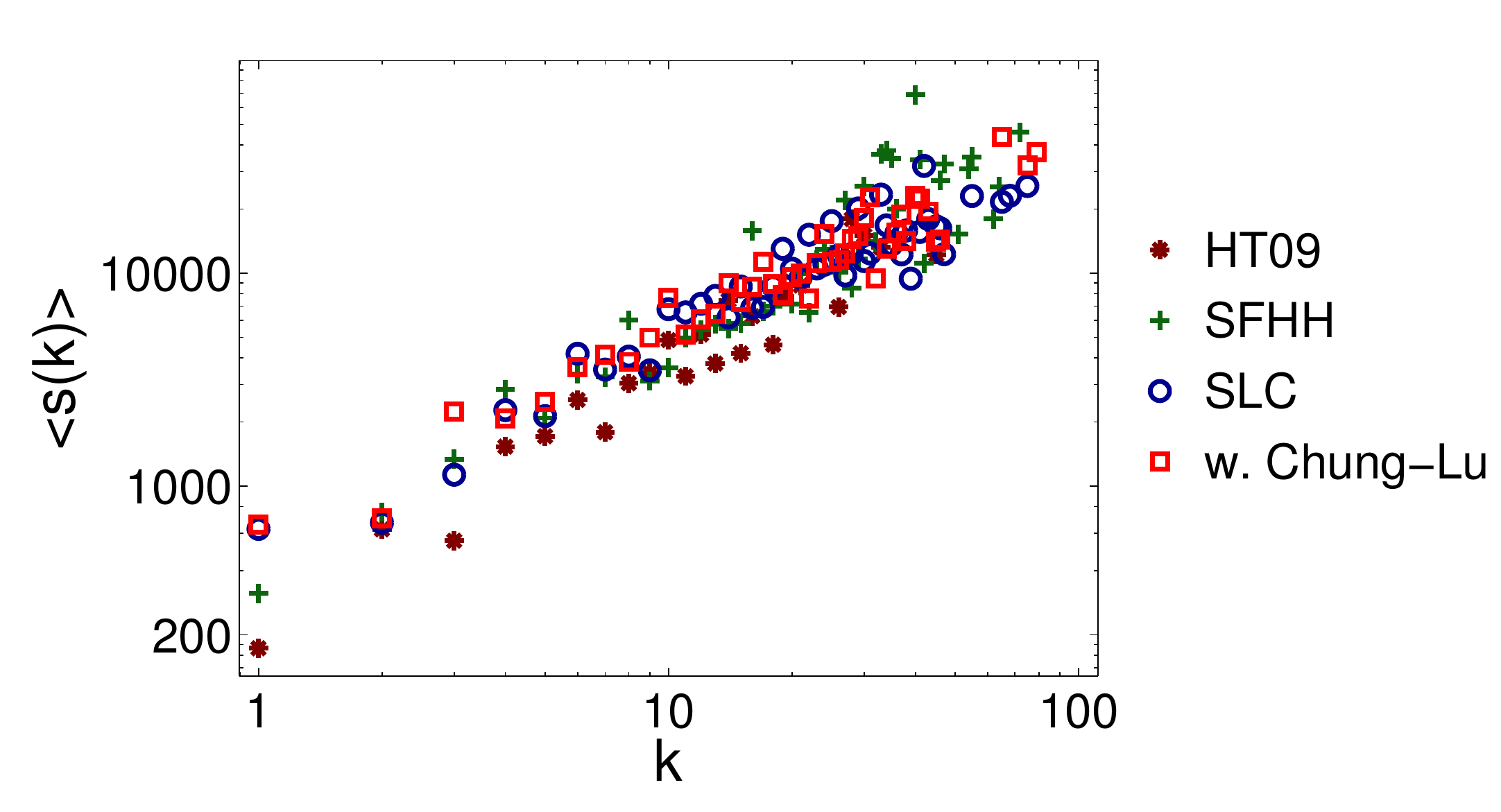}
\end{center}
\caption{(Color online) Average strength versus degree of nodes in three different
  scientific conferences (described in Section~\ref{ssec:4B}). 
  The squares represent the same quantity for a weighted Chung-Lu
  graph generated from the empirical distributions of the SLC
  contact network.}
\label{degree_strength_corr}
\end{figure}

As we are here interested in weighted networks, we introduce a weighted version of
this model that takes into account the fact that, in many real networks, weights and topology 
are not independent~\cite{Barrat2004}. 
This is in particular 
the case in the networks of face-to-face contacts considered in Appendix~\ref{sec:4} and Section~\ref{sec:5}, 
as illustrated in Fig.~\ref{degree_strength_corr}. Note that, depending on the data at hand, other models 
could be used to estimate $\delta^*$, $\chi^{2*}$ and $\sigma_u^*$. 
In our case of weighted networks with dependences between weights and topology, we propose the 
following variation to the classical Chung-Lu model.

We first compute the empirical distribution $P_k(w)$ of the weights of the links attached to nodes of degree $k$
from the real data of Section~\ref{sec:4}, for each degree $k$.
We then create a Chung-Lu graph with the same expected degree
sequence as the real data. For each node $i$ (of degree $k_i$) of this
Chung-Lu graph, we draw weights from the appropriate
distribution $P_{k_i}(w)$ and randomly allocate them to the links $i-j$ whose weights have not
yet been specified (if $i$ is linked to a node $\ell$ that has already been 
considered in the procedure, the weight of link $i-\ell$ has already been chosen by using
$P_{k_\ell}$ and it does not need to be computed again). In this way,
the weight sequence will be similar to the empirical graph's, if not exactly the same.
We thereby obtain a weighted Chung-Lu graph with the same expected degree sequence, the same strength-degree 
correlation and a similar weight sequence as the empirical graph of Section~\ref{sec:4}. 
Figure~\ref{degree_strength_corr} shows that the strength-degree correlation of such a weighted Chung-Lu graph
is indeed in agreement with the empirical data. Each Chung-Lu graph we generate can be seen as a topologically randomised 
version of the graph of contacts.
 
Note that this randomisation concerns the whole graph, and is in no way related to the one proposed
for the bootstrap samples. Hence, there is no impediment to use these weighted
Chung-Lu graphs as a controlled input for validating the statistical test discussed in Section~\ref{sec:2}.

\subsection{Validation of the bootstrap test}

\begin{figure}
\begin{center}
\includegraphics[width=0.5\textwidth]{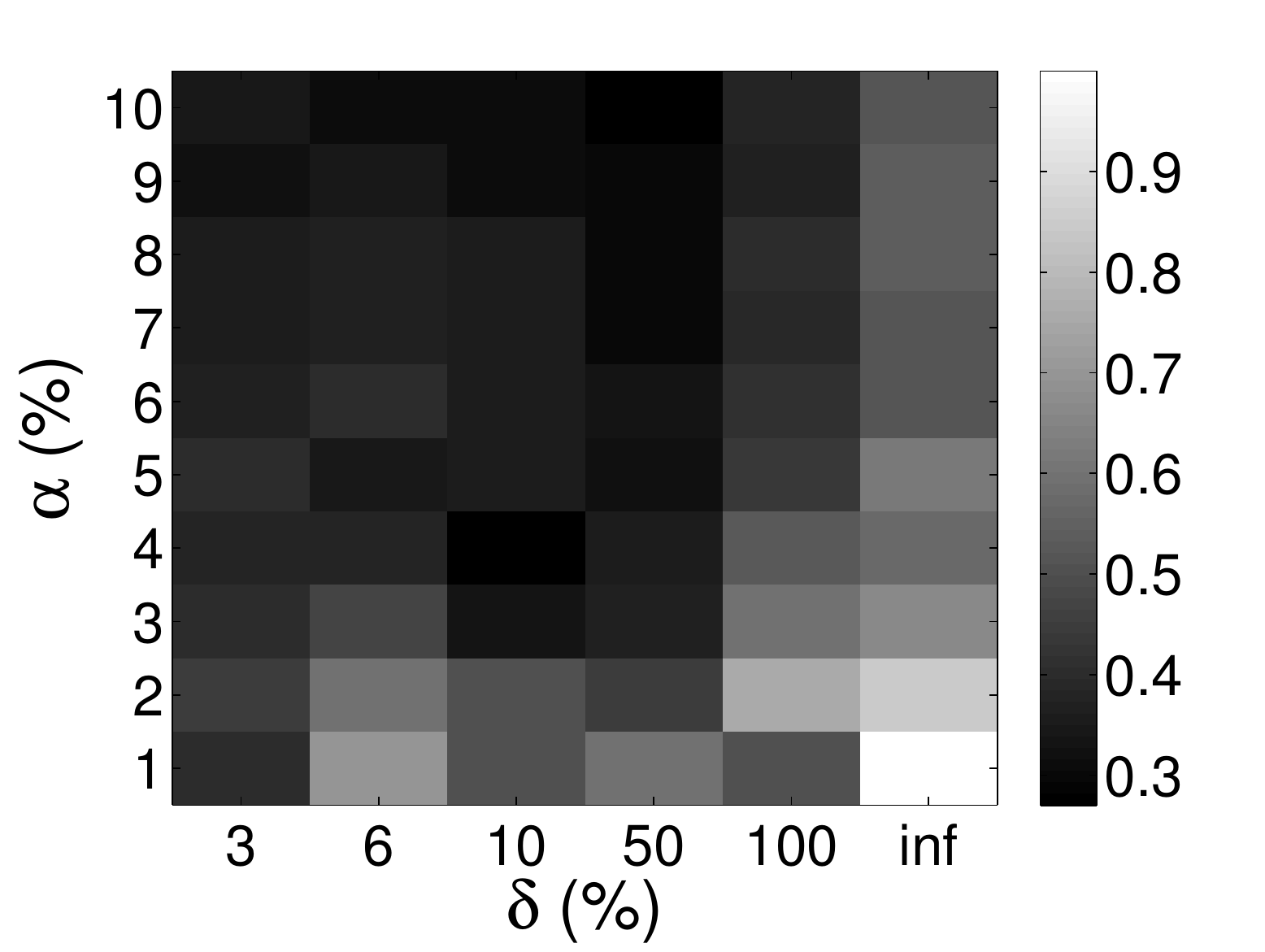}
\end{center}
\caption{\ADDED{Ratio of the obtained false alarm rate (probability of rejecting the Null 
Hypothesis when it is true) divided
by the maximum 
(pessimistic) significance level $\alpha$ of the test, in the case of the 
weighted Chung-Lu model, for the Null Hypothesis: ``same cardinality and same modularity''.
The prescribed significance level $\alpha$ acts as expected by the Bonferroni correction 
as a pessimistic bound to the true false alarm rate 
(hence, the ratio is always lower than 1). 
The results are displayed as function of $\alpha$ and $\delta$. It shows that the 
larger $\alpha$ is, the less tight the bound is.
When $\delta$ increases, the obtained false alarm rate becomes closer to $\alpha$.}
}
\label{CL_false_alarm} 
\end{figure}

\ADDED{A Monte-Carlo approach is used to validate the proposed method in the case of
weighted Chung-Lu graphs.
The goal is here to check the false alarm rate (rate at which 
normal groups are rejected) of the test.
For this purpose, we generate $1000$ instances of weighted Chung-Lu graphs.
The Null Hypothesis selected is ``having the same modularity and the same cardinality'' 
as $X^0$ as it is one of  the 
most representative Hypothesis to assess a group's behavior with respect to human social contacts.
The described method is applied to $1000$ random sub-groups (one in each of the thousand 
generated weighted Chung-Lu graphs), with different 
significance levels $\alpha$ (from $1\%$ to $10 \%$) and relaxation factors $\delta$ 
for the modularity constraint (from $0.03$ to $1.0$, with
an additional case $\delta = \infty$ that corresponds to no constraint on modularity). 
In general, a random group in a weighted Chung-Lu graph should be classified as normal, 
and should not be rejected by the test. This is what we verify here.

Figure~\ref{CL_false_alarm} shows the false alarm rate (i.e., the frequency of 
rejection of the Null Hypothesis) 
that is obtained in these simulations, divided by the prescribed significance 
level $\alpha$. For the test to be sound, this value
has to always be bounded by $1$. This is indeed the case. 
Moreover, the value is often much lower than $1$ (between $0.3$ and $0.6$).
This is a sign that the Bonferonni correction is pessimistic; it conducts 
us to reject more often the Null Hypothesis than
we should when the group under study is drawn according to the Null Hypothesis.
This is not an actual problem in the study as we are more interested in being 
certain that a group that is not rejected is indeed normal 
-- and this is the case.
}

\subsection{Controlling the  size of the bootstrap space and the power of the test}
\label{subs:CL_tradeoff}

\begin{figure}
\begin{center}
  \includegraphics[width=0.35\textwidth]{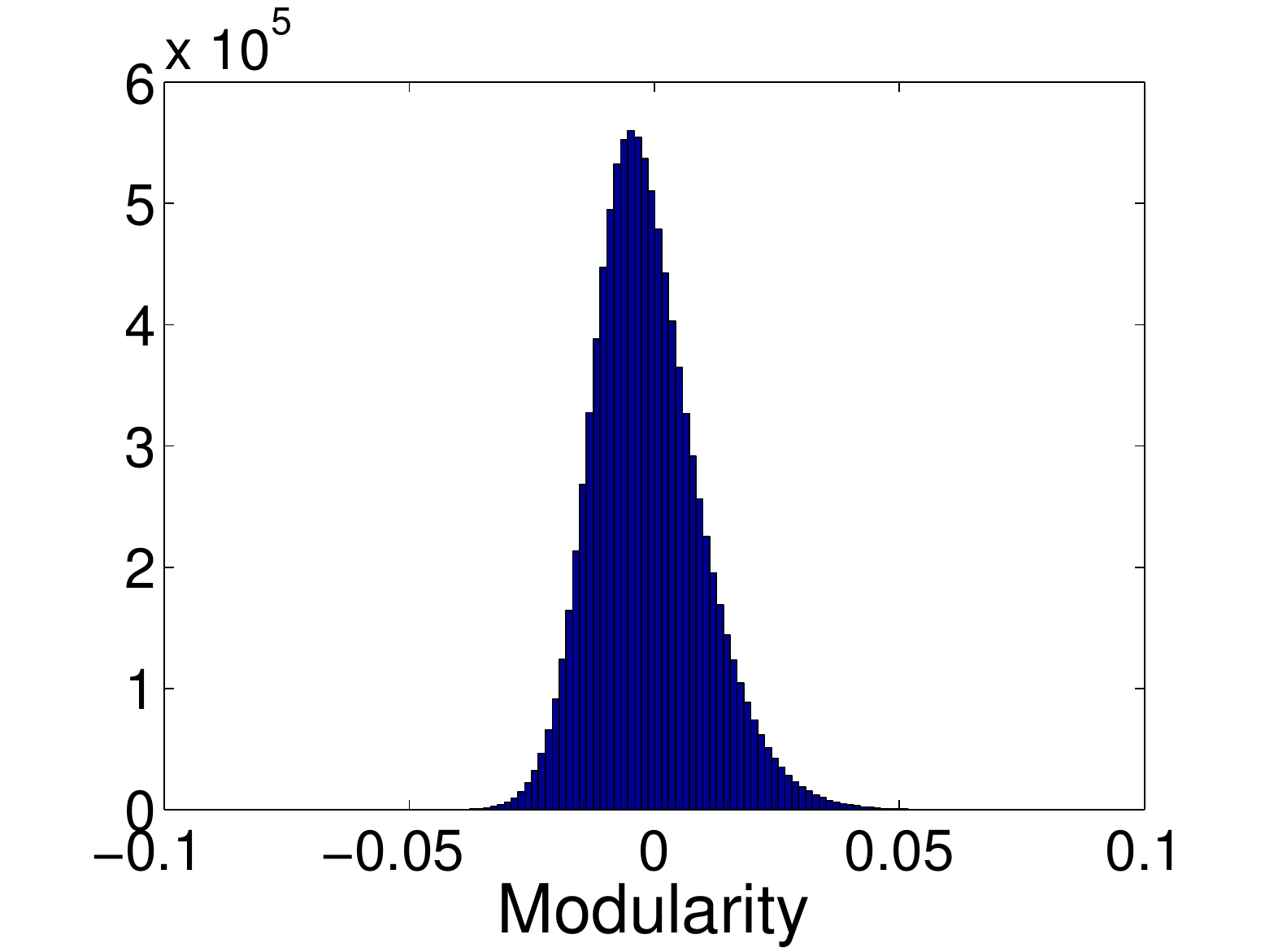}
 \end{center}

\caption{Histogram of the modularity of a sub-graph of $39$ nodes in
  a weighted Chung-Lu graph 
  }
\label{histo_MODU} 
\end{figure}

In order to derive thresholds $\delta^*$ and maximal values $\chi^{2*}$ and $\sigma_u^*$ 
for each given constraint, we take a different perspective and use a notion 
of rarity: when a value is in the bulk of a distribution it is considered as 
common enough and when it is in the extreme tails, it is considered as too rare. 
To illustrate the argumentation, we focus on the constraint ``same cardinality and 
same modularity'', where the strength of the ``same modularity'' constraint is tuned by 
$\delta$. Once more, we consider $1000$ weighted Chung-Lu graphs (computed with 
the empirical distributions of the data of Appendix~\ref{sec:4}), and, within each 
Chung-Lu graph, $10000$ random groups of 
cardinal $M=39$ (one of the group sizes studied later on). Figure~\ref{histo_MODU} 
shows the histogram of the modularity for the partition of the
graph in such a group and its complementary. Typical groups give a small modularity, 
the mode of the histogram being between $-0.03$ and $0.03$.
We define as rare groups those whose modularity are in the extreme tails of 
the distribution: either larger than its $10^6$ upper quantile or 
smaller than its $10^6$ lower quantile (these quantiles are reasonably estimated as we 
have $10^7$ samples in the distribution). The choice of these particular quantiles 
is somehow arbitrary but it can be easily changed for
the following study and does not influence the general approach. This gives us two 
modularity boundaries $Q_l^*=-0.050$ and $Q_u^*=0.076$ that separate 
common enough groups (having modularities in the bulk of the distribution)
from rare groups. \\

We take the point of view that the test should indicate that the Null Hypothesis is 
true for all common groups (output number 2 in the list of Section~\ref{subs:tradeoff}). 
Indeed, weighted Chung-Lu graphs are random and random groups in these graphs have 
in general no reason to be abnormal.
Now, consider a group $X^0$ with modularity around $-0.005$ (the peak of the 
distribution). For a given $\delta$, the simulated annealing procedure will 
draw bootstraps from this distribution: there is a high chance that the bootstrap set's 
modularities end up close to $-0.005$. Hence $X^0$ is compared to very similar 
groups even for high $\delta$: the test will always output 
$d=0, \chi^{2}<\chi^{2*}, \sigma_u<\sigma_u^*$ (apart maybe for extremely small $\delta$). 
But as one approaches the modularity boundaries $Q^*$, the test will start 
to misclassify $X^0$ as abnormal for a large enough $\delta$. Indeed, consider a group 
with a modularity close to or equal to one of the boundaries (for instance $Q_u^*$). For a too large 
$\delta$, the bootstrap set's modularities will still tend towards the small 
modularities that have a higher chance to be picked, and the test will be rejected. 
If we want all common groups with high modularity to be classified as normal, we need 
to have small enough $\delta$: this defines a first bound $\delta_u$ for $\delta$:
$\delta<\delta_u$. The same argumentation holds for common groups with low modularity 
(close to $Q_l^*$), giving another bound $\delta_l$ for $\delta$: $\delta<\delta_l$. Overall, 
we thus obtain an upper bound for $\delta^*$: $\mbox{min}(\delta_l,\delta_u)$. In order to
decide where in the range $[0, \mbox{min}(\delta_l,\delta_u)]$ we 
should  choose $\delta^*$,  the trade-off discussion of 
Section~\ref{subs:tradeoff} between precision of the Null Hypothesis and 
power of the test still holds. We give priority to the power of the test and choose the maximum 
possible value of $\delta$, equal to the upper bound: $\delta^*=\mbox{min}(\delta_l,\delta_u)$. In practice, as 
shown in Table~\ref{tab:stars}, $\delta^*$ is between $5\%$ and $15\%$: 
the Null Hypotheses are still reasonably precise.\\

\begin{figure}
\begin{center}
\includegraphics[width=0.45\textwidth]{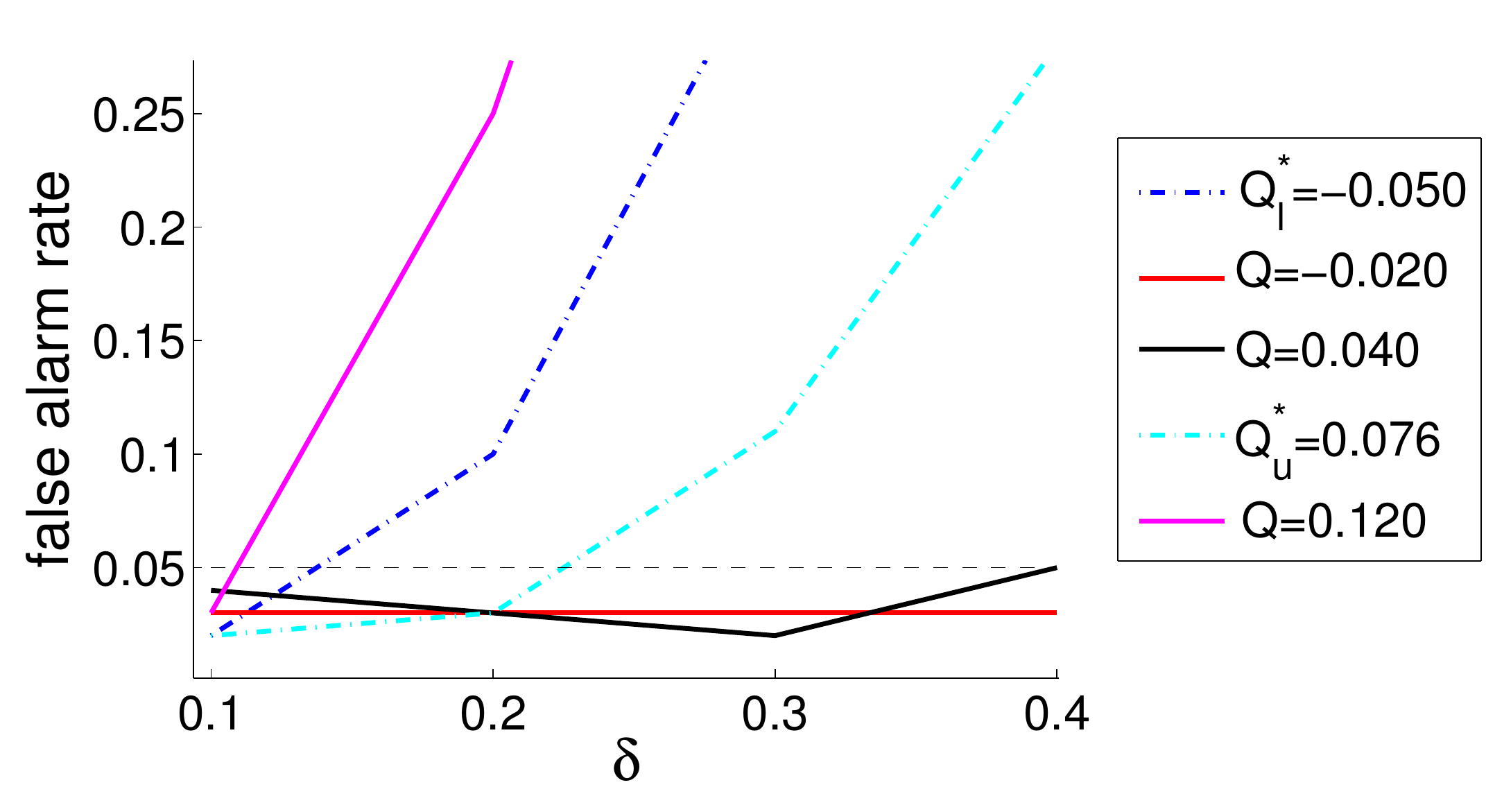}
\end{center}
\caption{
\ADDED{(Color online) False alarm rate as function of $\delta$ for 
groups of 39 nodes of varying modularities (describing the rarity of the groups) 
in weighted Chung-Lu graphs.
Groups with modularity $Q$ outside the interval $([Q_l^*=-0.05, Q_u^*=0.076])$  
are rare, 
and the others are considered common enough. 
To keep the false alarm rate under the expected significance level $\alpha$ 
(here equal to $5\%$ and 
represented by the horizontal dashed line)
for all common groups, maximum values $\delta_l^*=0.15$ and $\delta_u^*=0.20$ 
are read on the plot. 
This in turn gives a general threshold $\delta^*=\mbox{min}(\delta_l^*, \delta_u^*)=0.15$.}
}
\label{fig:false_alarm_vs_modul} 
\end{figure}

\begin{figure}
\begin{center}
\begin{minipage}{0.45\linewidth}
\begin{center}
 \includegraphics[width=\textwidth]{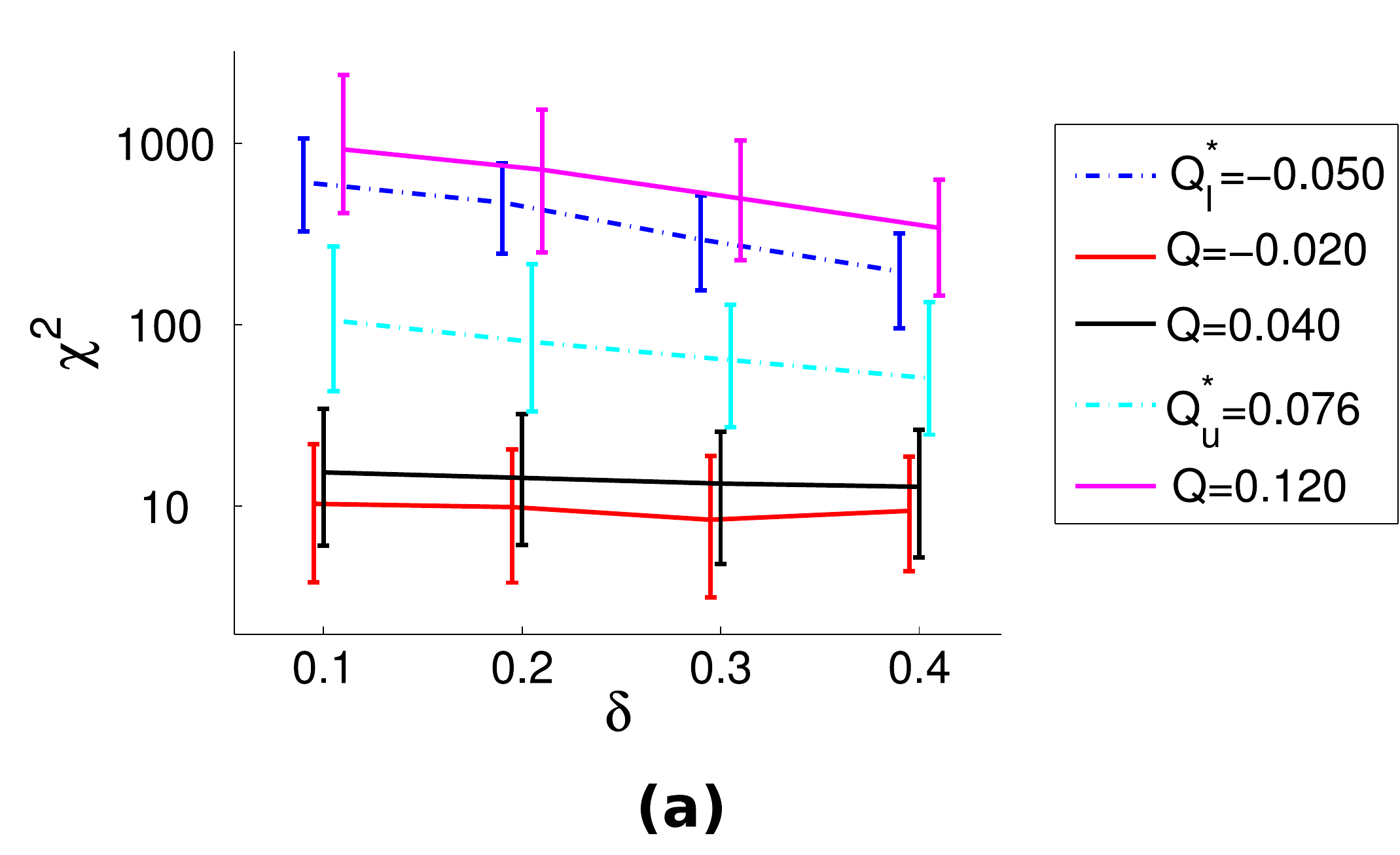}
\end{center}
\end{minipage}
\hfill
\begin{minipage}{0.45\linewidth}
\begin{center}
\includegraphics[width=\textwidth]{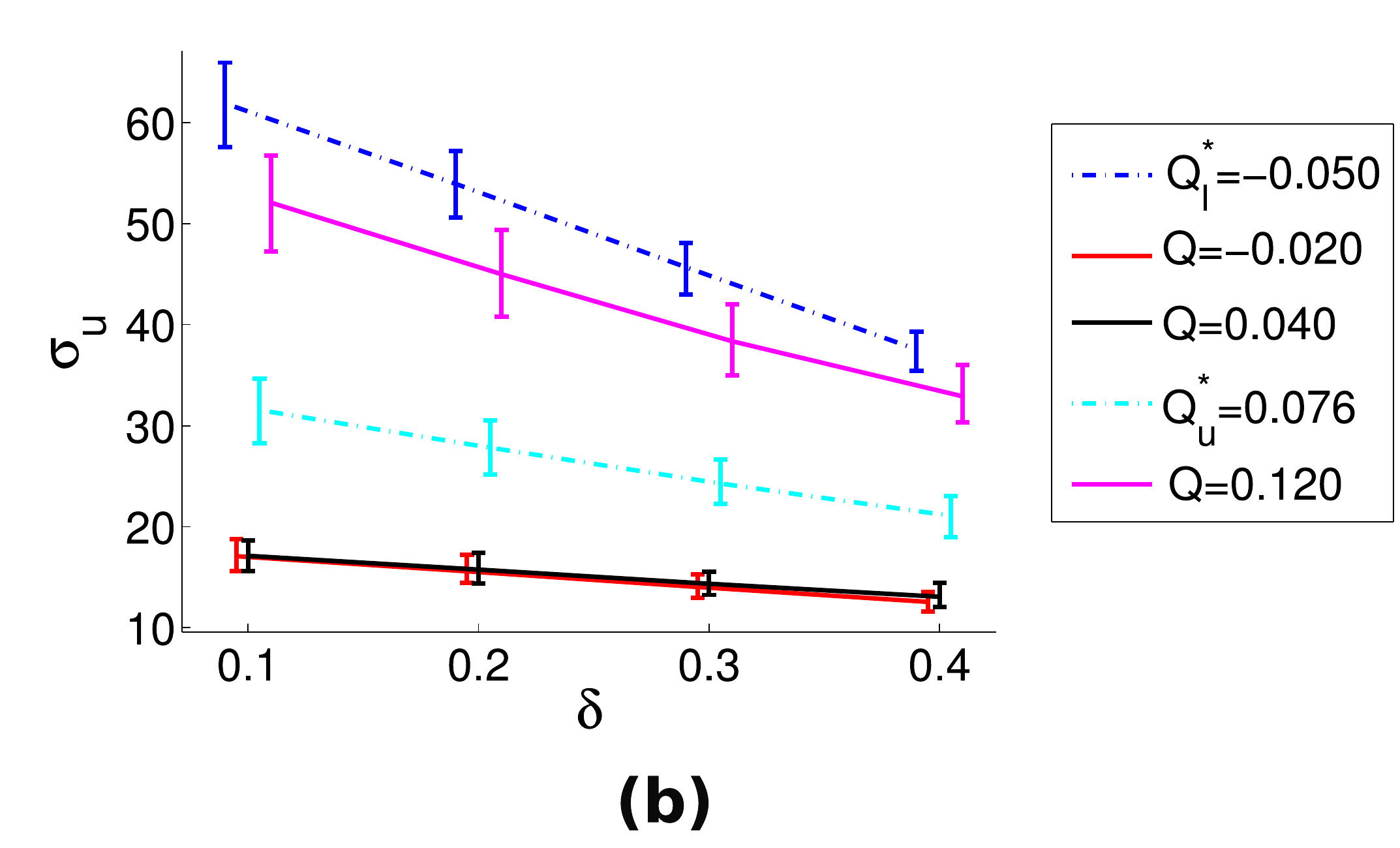}
\end{center}
\end{minipage}
\end{center}
\vspace{-0.5cm}
\caption{(Color online) 
\ADDED{$\chi^2$ (a) and  $\sigma_u$ (b) as a function of $\delta$ for
groups of varying modularities (describing the rarity of the groups)  in weighted Chung-Lu 
graphs.
As the test is designed to be valid for all common enough groups (i.e., with modularity 
$Q \in[Q_l^*=-0.05, Q_u^*=0.076]$) if
one uses $\delta^*=0.15$, the maximum values $\sigma_u^*$ and $\chi^{2*}$ that should be 
accepted for  
$\sigma_u$ and $\chi^2$ are read from the curves. 
This procedure gives roughly $\chi^{2*}=950$ and $\sigma_u^*=60$.}
}
\label{fig:chi2sigma_CL} 
\end{figure}

Figure~\ref{fig:false_alarm_vs_modul} shows the false alarm rate obtained in
simulations for weighted Chung-Lu graphs as a function of $\delta$ 
for groups of 39 nodes of varying modularities
(both common and rare). As expected, for very common groups (like groups with $Q=0.02$), 
the false alarm rate is constant with respect to $\delta$. On the contrary, the more 
we consider groups 
close to the boundaries $Q_l^*$ and $Q_u^*$, the faster the rate increases with $\delta$.
The simulations were conducted on 100 different Chung-Lu graphs.
To keep the false alarm rate under the expected significance level $\alpha$ 
(here equal to $5\%$)
for all common groups, a maximum value $\delta^*$ for $\delta$ is obtained from this Figure. 
For the lower quantile $Q_l^*$ one reads $\delta_l=0.15$ and for the upper quantile $Q_u^*$
we obtain $\delta_u=0.20$. The general bound for the modularity constraint 
for groups of $39$ nodes is then $\delta^*=\mbox{min}(\delta_l, \delta_u)=0.15$. 
In fact, if one chooses $\delta<\delta^*$, all normal common groups will be correctly 
classified (with a tolerance of $\alpha=5\%$).\\

The next step is to find thresholds in the acceptable values for $\sigma_u$ and $\chi^2$. 
On Figure~\ref{fig:chi2sigma_CL}, these indicators are displayed as a function of $\delta$ 
for groups of $39$ nodes of varying modularities. 
As expected, $\chi^2$ and $\sigma_u$ increase monotonically with the
rarity of the considered groups (for a fixed value of $\delta$).
As we argued that the test is designed to classify all common groups as normal
if one uses $\delta = \delta^*$, the maximum values $\sigma_u^*$ and $\chi^{2*}$ 
that have to be tolerated for $\sigma_u$ and $\chi^2$ are read from the curves as 
the maximum expected values for these common groups (we use the quantile at $95\%$ 
to fix a reasonable maximum value). One reads the approximate values $\chi^{2*} (\delta^*;Q=Q_l^*)=950$, 
$\sigma_u^* (\delta^*;Q=Q_l^*)=60$ and $\chi^{2*}(\delta^*;Q=Q_u^*)=250$, 
$\sigma_u^*(\delta^*;Q=Q_u^*)=30$. The general thresholds are therefore: 
$\chi^{2*}=\mbox{max}(\chi^{2*}(\delta^*;Q=Q_l^*),\chi^{2*}(\delta^*;Q=Q_u^*))=950$ and 
$\sigma_u^*=\mbox{max}(\sigma_u^*(\delta^*;Q=Q_l^*),\sigma_u^*(\delta^*;Q=Q_u^*))=60$.\\

The final conclusion of this validation procedure is that, once one has decided upon $\alpha$, 
a type of constraint, the cardinality of groups of interest, and a criterion to decide
what common enough (i.e., not too rare) means for groups, it is possible to quantify the 
bootstrap approach presented here and to propose a value $\delta^*$ and thresholds 
$\sigma_u^*$ and $\chi^{2*}$. 
We show $(\delta^*, \sigma_u^*, \chi^{2*})$ for all the different constraints and 
all the different cardinalities used in Section~\ref{sec:5} in Table~\ref{tab:stars}-a. 
In Appendix~\ref{sec:4} and Section~\ref{sec:5}, we will compare the output of the test for four different 
groups with different Null Hypotheses. To be able to compare properly, we will use a 
unique value of $\delta^*$ for each Null Hypothesis. Therefore, out of the 
four $\delta^*$ proposed (one for each cardinality), we keep the minimum one and 
obtain the equivalent $\sigma_u^*$ and $\chi^{2*}$ for each cardinality. We 
sum up these values in Table~\ref{tab:stars}-b.

For the study presented in this paper, we have therefore used these values of $\delta^*$, $\chi^{2*}$ and 
$\sigma_u^*$ obtained with the weighted Chung-Lu graphs used as surrogates of social networks.

\begin{table}
\ADDED{
\begin{center}
 \begin{tabular}{c||c|c|c|c}
 Null Hypothesis & $M=39$  & $M=73$ & $M=99$ & $M=106$\\\hline\hline
$Q_X$ constraint& $(0.15, 950, 60)$  & $(0.15, 110, 45)$ & $(0.15, 40, 35)$& $(0.15, 40, 30)$\\\hline
$N_{XX}$ constraint& $(0.15, 3500, 85)$  & $(0.05, 2700, 115)$ & $(0.05, 2600, 125)$& $(0.05, 2200, 125)$\\\hline
$T_{XX}$ constraint&  $(0.15, 3600, 80)$  & $(0.1, 2200, 100)$ & $(0.05, 3700, 135)$& $(0.05, 2700, 135)$\\
\end{tabular}

\vspace{0.2cm}

a)
\vspace{0.5cm}

 \begin{tabular}{c||c|c|c|c}
 Null Hypothesis & $M=39$  & $M=73$ & $M=99$ & $M=106$\\\hline\hline
$Q_X$ constraint with $\delta^*=0.15$& $(950, 60)$  & $(110, 45)$ & $(40, 35)$& $(40, 30)$\\\hline
$N_{XX}$ constraint with $\delta^*=0.05$& $(5200, 95)$  & $(2700, 115)$ & $(2600, 125)$& $(2200, 125)$\\\hline
$T_{XX}$ constraint with $\delta^*=0.05$&  $(4300, 90)$  & $(3700, 120)$ & $(3700, 135)$& $(2700, 135)$\\
\end{tabular}

\vspace{0.2cm}

b)
\end{center}
 \caption{ a) $(\delta^*, \sigma_u^*, \chi^{2*})$ for different constraints and 
cardinalities. $(\delta^*, \sigma_u^*, \chi^{2*})$ for the 
 $Q_X$ constraint and $M=39$ are read from Figures~\ref{fig:false_alarm_vs_modul} 
 and \ref{fig:chi2sigma_CL} as explained in the text. Figures used to obtain 
 the values for other constraints and other cardinalities are not shown. 
 b) For each constraint, we decide to keep a unique $\delta^*$: the minimum 
 of the four $\delta^*$ (one for each cardinality). We show here the 
 corresponding $(\sigma_u^*, \chi^{2*})$ for all cardinalities.
}
 \label{tab:stars}}
\end{table}

\section{Presentation of the Dataset of Two Co-located conferences}
\label{sec:4}

\subsection{Data and pre-processing}
\label{ssec:4A}

\begin{figure}[h!]
\includegraphics[width=0.8\textwidth]{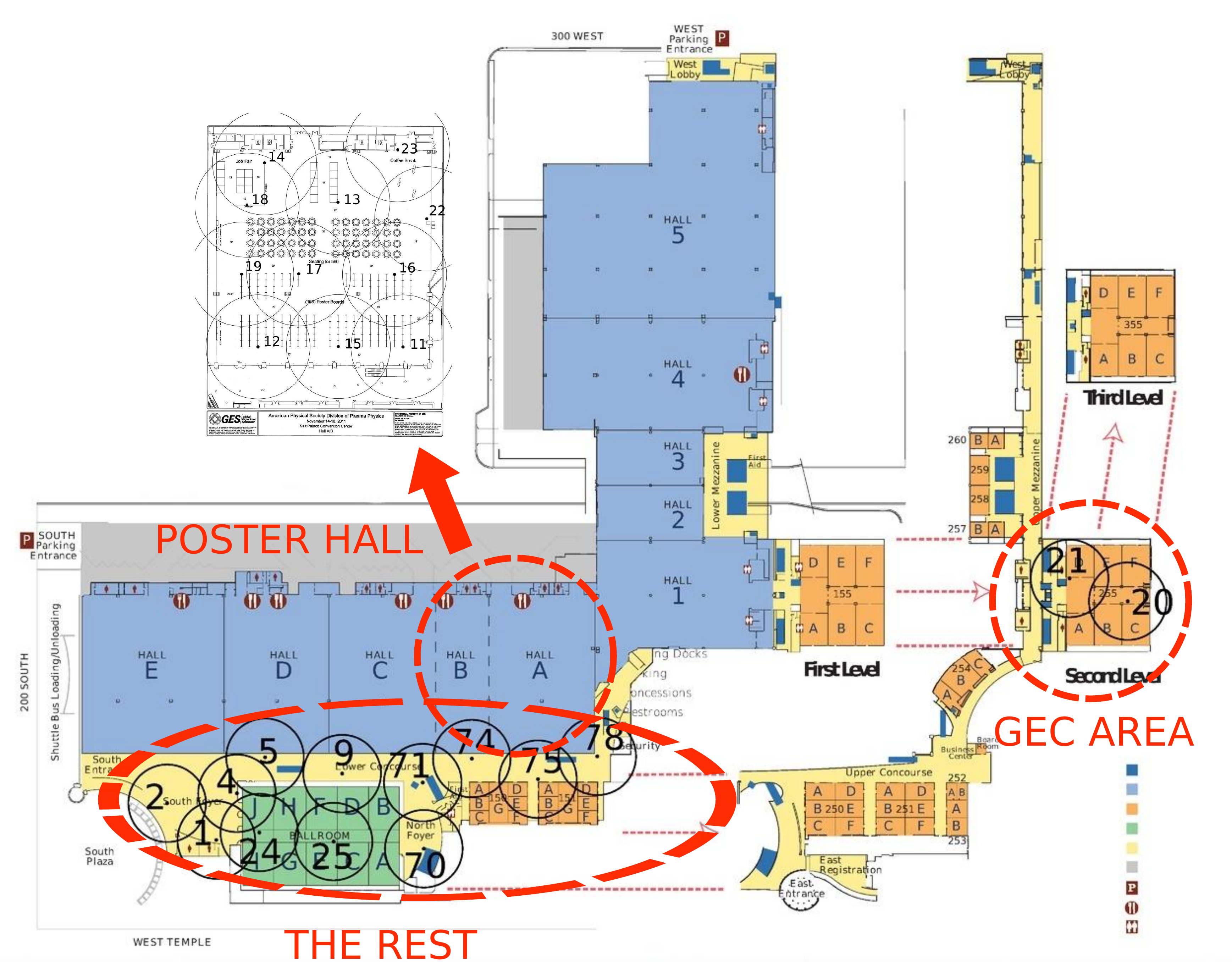}
\caption{General map of the conference venue with the three different general areas. 
Each black circle corresponds to one of the 25 RFID readers used to measure the social interactions. 
The GEC area is isolated: it is 500 meters away from the Poster Hall.}
\label{plans}
\end{figure}

The data was collected in Salt Lake City (SLC) in November 2011 during two 
co-located scientific conferences lasting five days, using the SocioPatterns sensing
infrastructure~\cite{sociopatterns,cattuto2010dynamics} to measure 
face-to-face proximity of individuals.  
The conferences  were jointly organised by the DPP
(Division of Plasma Physics) of the American Physical Society and the GEC (Gaseous Electronics Conference).
Figure~\ref{plans} shows the map of the conference venue in Salt Lake
City.

Out of the 2081 participants of the conference, 320 agreed to
participate in our study: 281 from DPP and 39 from GEC.  The
participation was on a voluntary basis so that there was no specific
sampling scheme.  The face-to-face proximity of the participants was
measured using the SocioPatterns sensing
infrastructure~\cite{sociopatterns,cattuto2010dynamics} based on
unobstrusive active RFID tags that can be embedded in conference
badges. Two tags exchange radio packets only if the individuals
wearing them face each other (the human body acts as a shield at the
frequency and power of the radio packets) within a distance of 1 to
1.5 meters. The detected proximity relations are reported by the tags
to RFID readers installed in the environment. At the end of the
conference, the raw data consists of a log of all the recorded
contacts. The log is a sequence of lines $(t, r, i, j)$ where $t$ is
the time at which reader $r$ received the information that the
individuals wearing tags $i$ and $j$ were in close face-to-face
proximity (``in contact'').
Given the operating parameters of the tags, proximity of two
individuals wearing the RFID badges can be assessed with a probability
in excess of $99\%$ over an interval of $20$ seconds
\cite{cattuto2010dynamics}, which is a fine enough time scale to
resolve human mobility and proximity at social gatherings.
We therefore aggregate the raw data over time windows of $20$ seconds:
we partition the five days of data gathering into $20$ second periods, and we associate
to each of these periods $t$  the adjacency matrix $A^{t}$ representing the aggregated graph over the 
20 seconds: $A_{ij}^{t}=1$ if and only if vertices $i$ and $j$ have exchanged at least one radio packet
during the time window $t$, otherwise $A_{ij}^{t}=0$. 

Overall, the data define a temporal contact network in which nodes represent individuals, and 
a link between two nodes at time $t$ denotes the fact that the corresponding individuals
are in face-to-face proximity. The temporal network can moreover be aggregated over the total duration
of the conference, defining a weighted contact network where each node is an individual and where 
the weight of a link between two individuals gives the cumulated time they have spent in face-to-face interaction 
during the conference.

\subsection{Distributions of contact durations}
\label{ssec:4B}
\begin{figure}
\begin{center}
\begin{minipage}{0.45\linewidth}
\begin{center}
 \includegraphics[width=\textwidth]{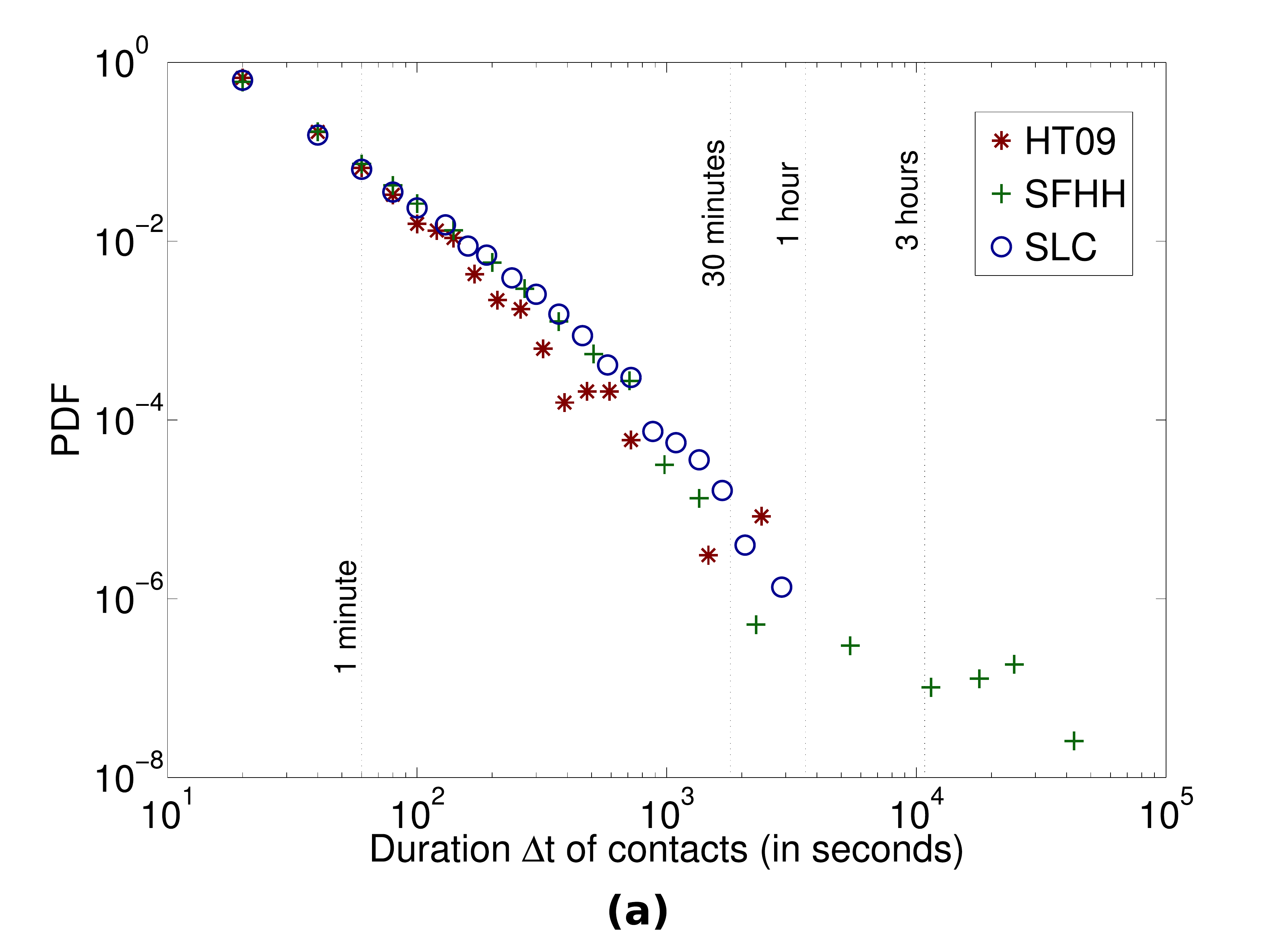}
\end{center}
\end{minipage}
\hfill
\begin{minipage}{0.45\linewidth}
\begin{center}
\includegraphics[width=\textwidth]{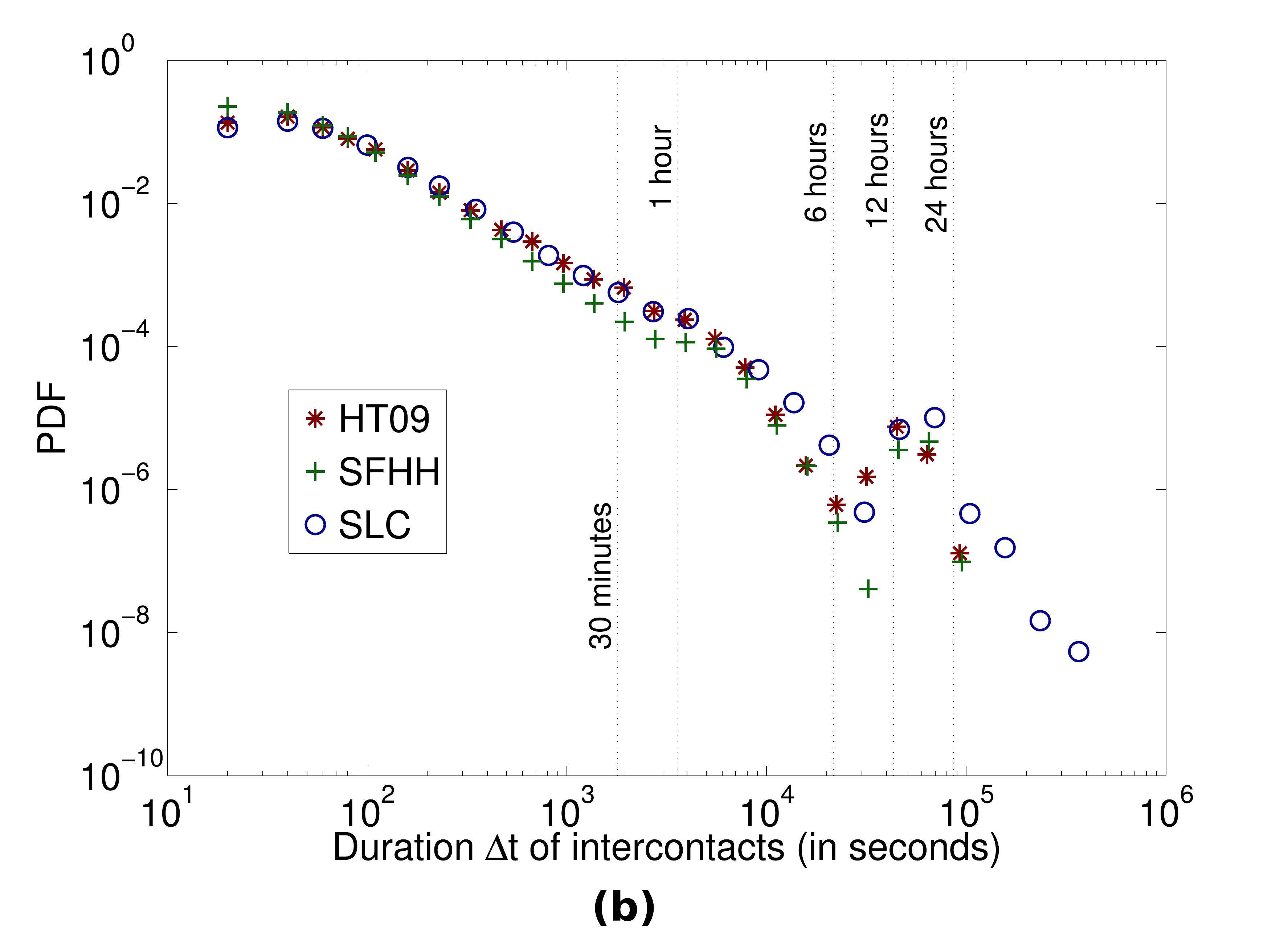}
\end{center}
\end{minipage}
\begin{minipage}{0.48\linewidth}
\begin{center}
 \includegraphics[width=\textwidth]{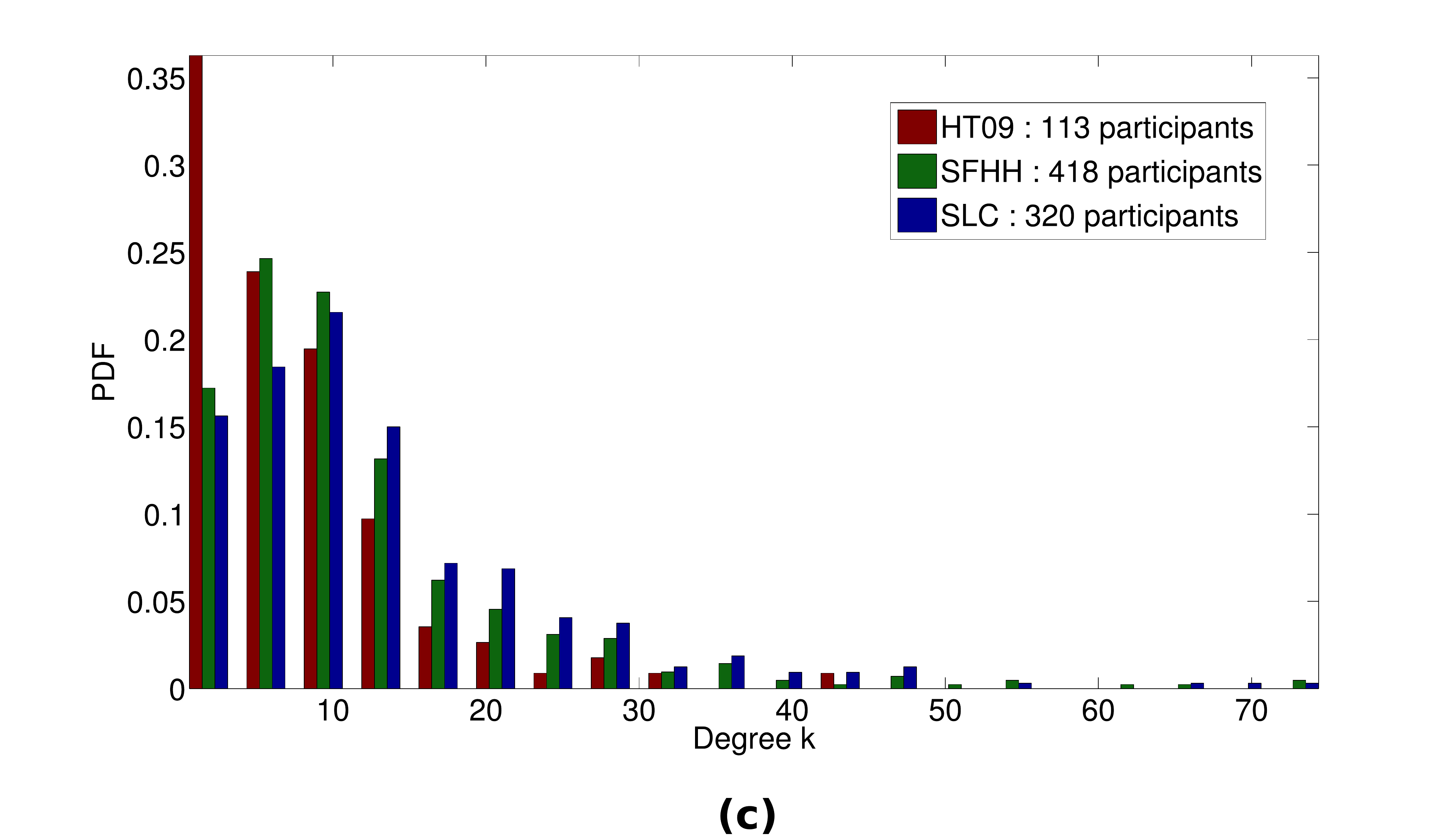}
\end{center}
\end{minipage}
\hfill
\begin{minipage}{0.45\linewidth}
\begin{center}
\includegraphics[width=\textwidth]{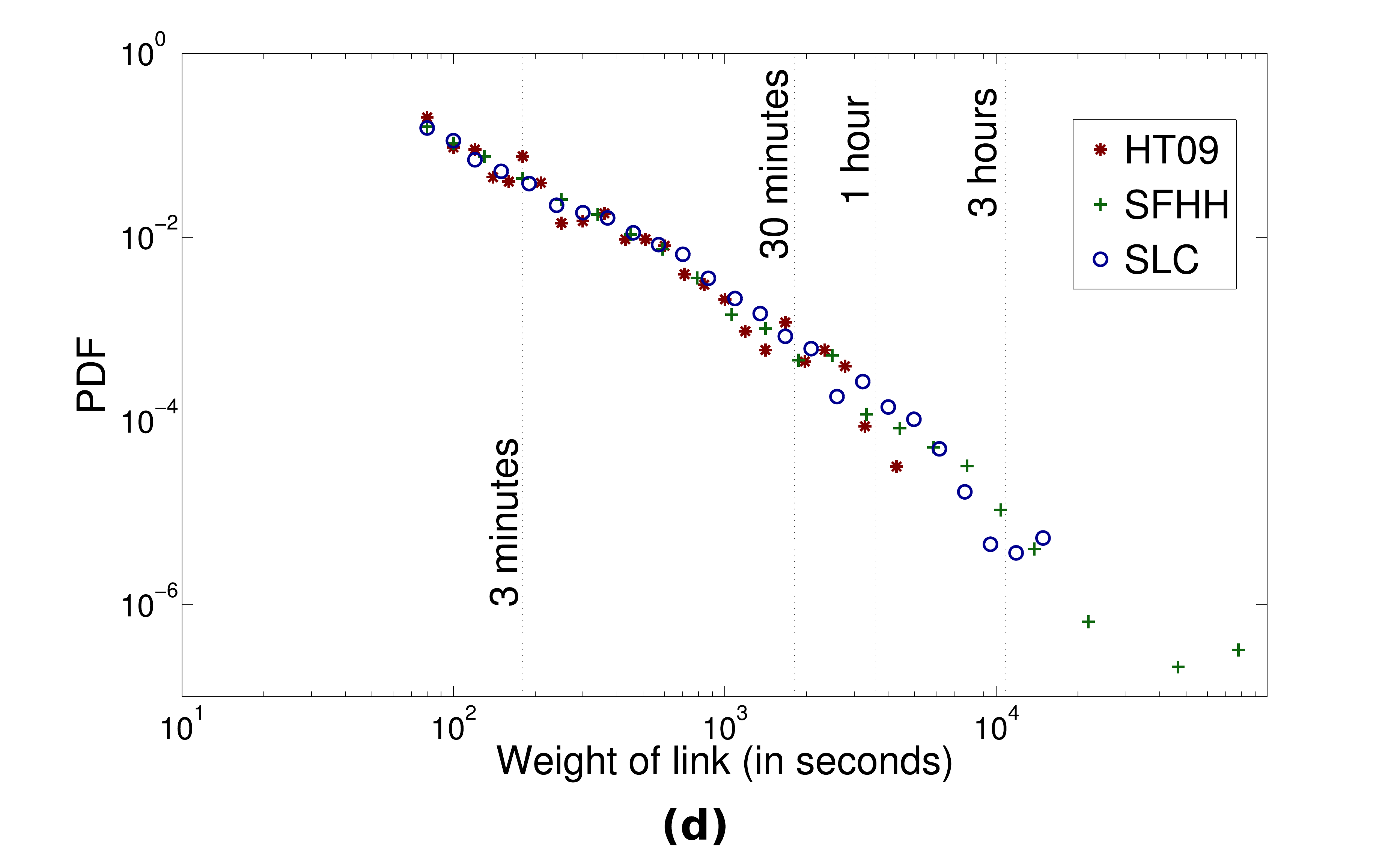}
\end{center}
\end{minipage}
\end{center}
\vspace{-0.5cm}
\caption{(Color online) (a) Comparison of the distribution of the durations of contacts for three different datasets.
(b) Distribution of the duration of intercontact times. An intercontact interval is 
defined as the interval, for each node, between the starting times of two successive contacts.
(c) Distribution of the degrees in the aggregated network of the three conferences. The degree of a participant corresponds 
to the total number of participants with whom s/he has been in contact during the conference.
(d) Distribution of link weights. The weight of a link between two nodes gives the total time in contact of the corresponding participants.
}
\label{contact}
\end{figure}

We first compare briefly the gathered data with other datasets collected in similar contexts using the same infrastructure. 
We define a contact between two tags $i$ and $j$ as an unbroken subsequence of $1$'s within the sequence $\{A_{ij}^{t}\}$. 
Its duration is the length of this subsequence. 
Table~\ref{tab:tablecomp} presents basic statistics of the present data, together with the ones collected during
 the 2009 ACM HyperText conference (HT09)~\cite{isella2011s} and during 
a congress of the \textit{Soci\'et\'e Fran\c{c}aise d'Hygi\`ene Hospitali\`ere} 
(SFHH)~\cite{cattuto2010dynamics}. Note that the sum of the total number of contacts (and the total 
time of contact) within DPP and within GEC does not exactly account for the interactions for the conference taken as a whole (ALL), due
to the interactions between 
DPP and GEC. The SLC data contain a relatively small number of contacts, in comparison with 
the other conferences, taking into account the number of participants and the duration: 
this is due to the small sampling rate of the total population of the SLC conferences. 
Figure~\ref{contact} however shows that various statistical properties of the contact networks, 
such as the distribution of the duration of contacts,  the distribution of degrees, of the  
inter-contact times or of the weights of the links, 
 are however very similar in the three contexts.
This confirms the robustness of the main statistical properties of the networks of face-to-face contacts
between individuals observed in previous works~\cite{isella2011s,eagle2006reality,hui2005pocket}. 

\begin{table}
\begin{center}
 \begin{tabular}{l|c|c|c|c|c|}
	    \multicolumn{1}{c|}{} & \multicolumn{1}{c|}{HTT09} & \multicolumn{1}{c|}{SFHH} & \multicolumn{3}{c|}{SLC} \\
	    \cline{4-6}
	    \multicolumn{1}{c|}{ } &\multicolumn{1}{|c|}{ } &\multicolumn{1}{|c|}{ } & GEC & DPP & ALL \\
	    \hline
	    No. of tags & 113 & 418 & 39 &  281 &320 \\\hline
	    sample rate & 75\% & 33\% & 12\% & 16\% & 15\% \\\hline
	    No. of days & 2 & 2 & \multicolumn{3}{c|}{5} \\\hline
	    No. of contacts & 9582 & 27434 & 1189 & 21519 & 23920 \\\hline
	    Tot. time of contact (hours)& 102 & 414 & 18 & 306 & 339 \\
\end{tabular}
\end{center}
\caption{Basic statistics concerning the datasets collected in three different scientific conferences.}
\label{tab:tablecomp}
\end{table}

\begin{figure}
\begin{center}
\includegraphics[width=0.5\textwidth]{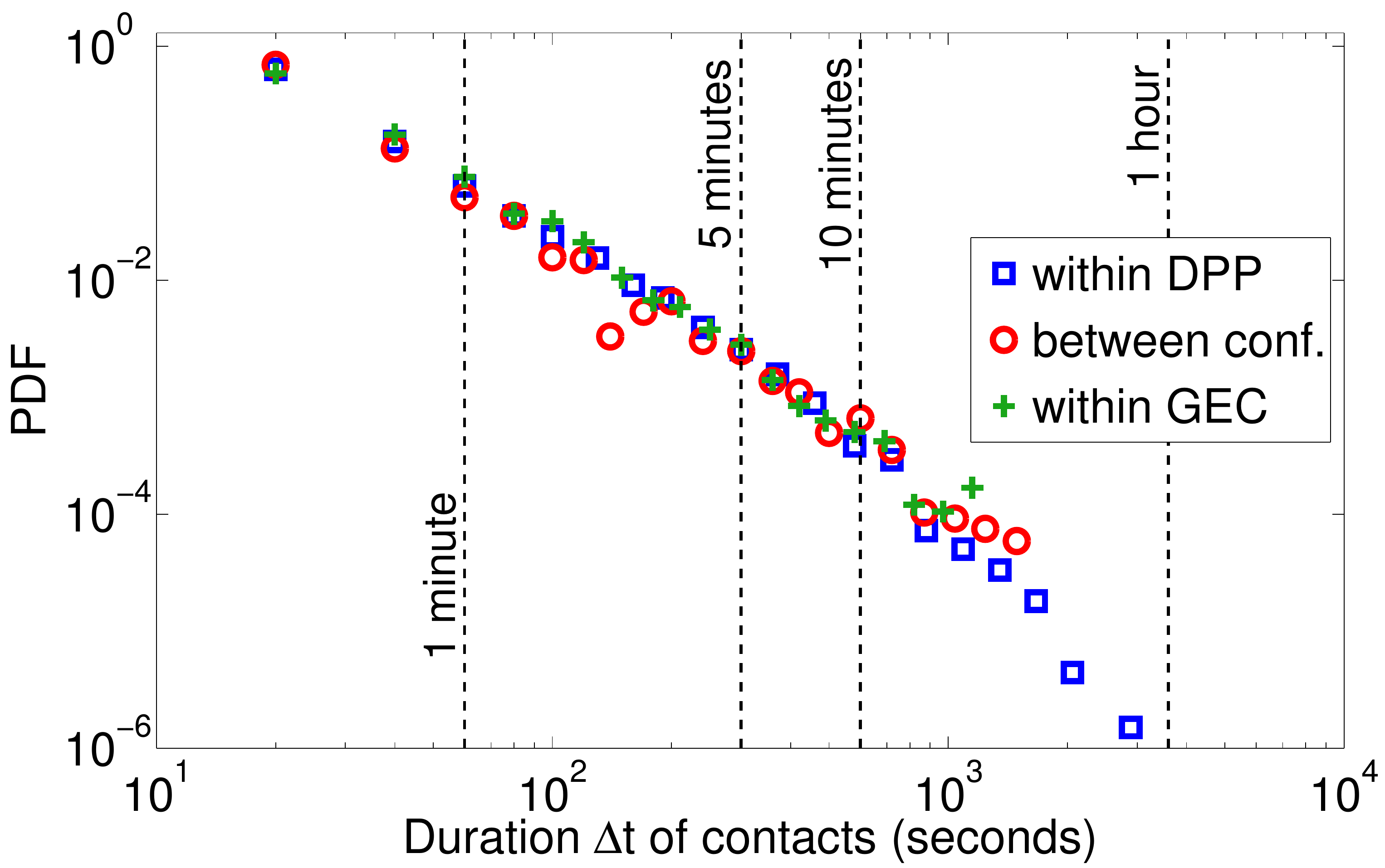} 
\end{center}
\caption{(Color online) Cumulative distributions of the durations of contacts within the DPP conference, within the
 GEC conference and between both conferences of the SLC dataset.}
\label{contactb}
\end{figure}

In the present dataset, we can distinguish three categories of contacts: 
within DPP, within GEC, and  between both groups. Figure~\ref{contactb} 
shows that even 
though the number of contacts is much larger within DPP than within GEC (see 
Table~\ref{tab:tablecomp}), 
the corresponding duration distributions collapse remarkably well upon one another.
Hence, we do not observe any difference in the statistical behavior of the three categories of contacts. 
Let us also note that we are not interested here in modeling these distributions (for instance by power law
or log-normal functional forms), as the method we will use is data-driven. It is however of interest to
remark that the broad shape of the distributions implies that parametric statistical methods would be 
hard to implement, and that data-driven statistical methods are expected to be more adequate.

\subsection{Distributions of the durations of contacts taking place in different areas}

The conference venue is spatially heterogeneous, with in particular
three broadly defined areas: \emph{the GEC Area} where the GEC
registration and coffee breaks took place; \emph{the Poster Hall},
where the poster sessions of both conferences took place; and
\emph{the Rest}, which includes the DPP registration desk, two coffee
break areas, and corridors linking different parts of the building.
The GEC Area was situated $500$ meters from the Poster Hall (maps are
shown in Annex B). It therefore took time and energy to walk from
one area to another, which was an obstacle to interactions between both groups.
As the measuring infrastructure allows us
to identify the area in which each reported contact took place, it is 
interesting to investigate if differences exist between the three types of contacts defined above
when the spatial information is taken into account.

\begin{figure}
\begin{center}
\begin{minipage}{0.32\linewidth}
\includegraphics[width=\textwidth]{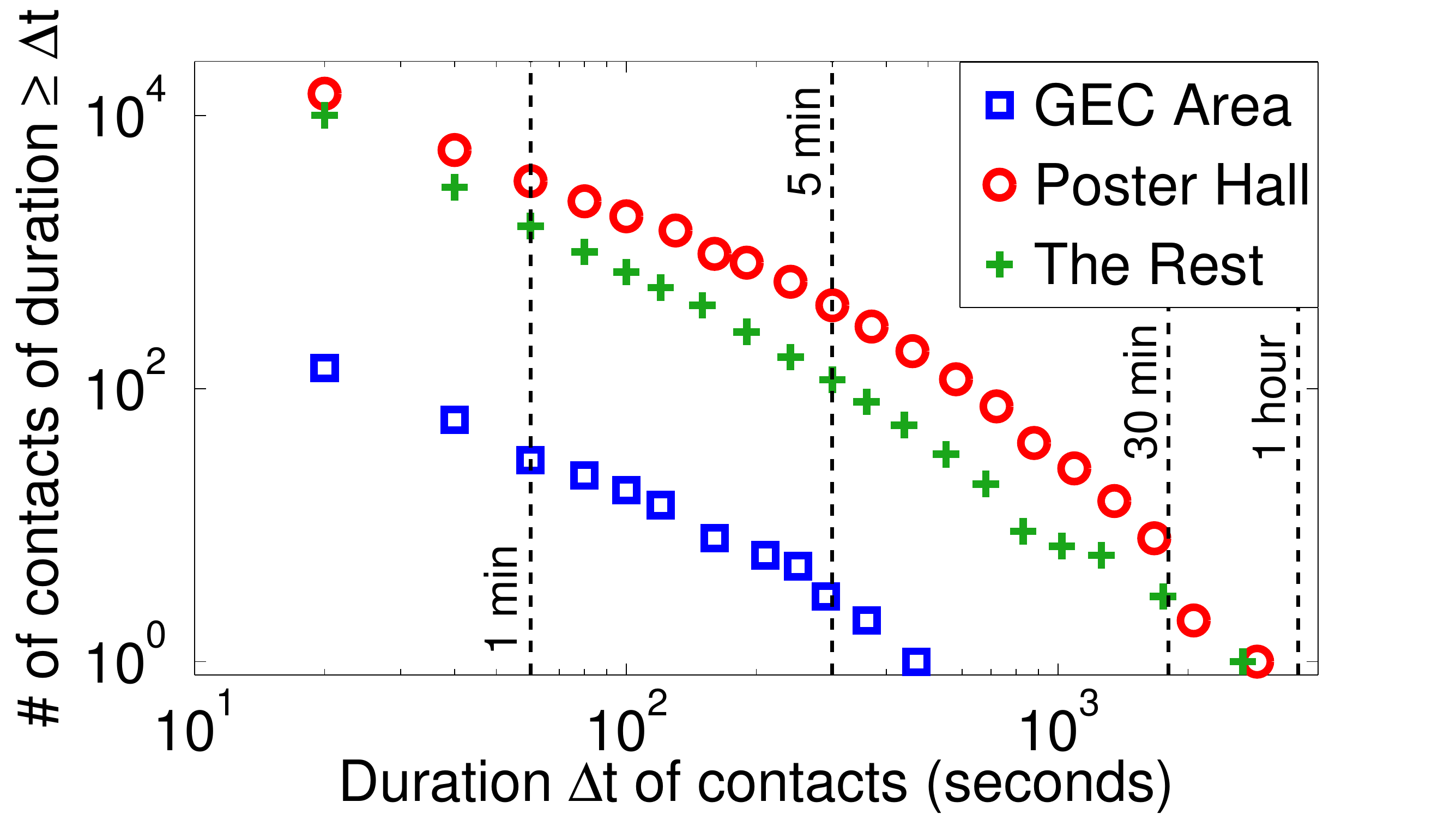} 
\end{minipage}
\hfill\begin{minipage}{0.32\linewidth}
\includegraphics[width=\textwidth]{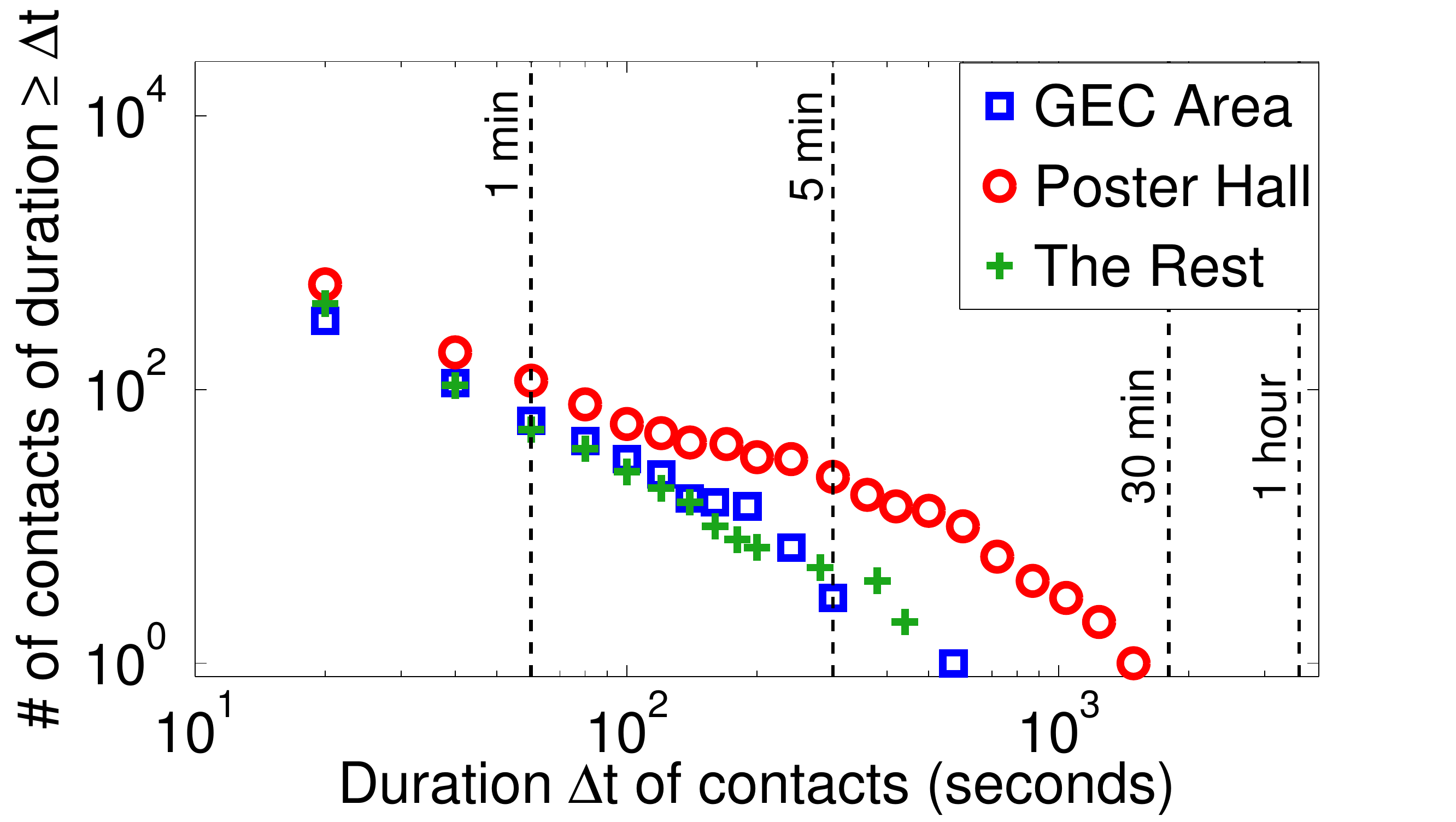} 
\end{minipage}
\hfill\begin{minipage}{0.32\linewidth}
\includegraphics[width=\textwidth]{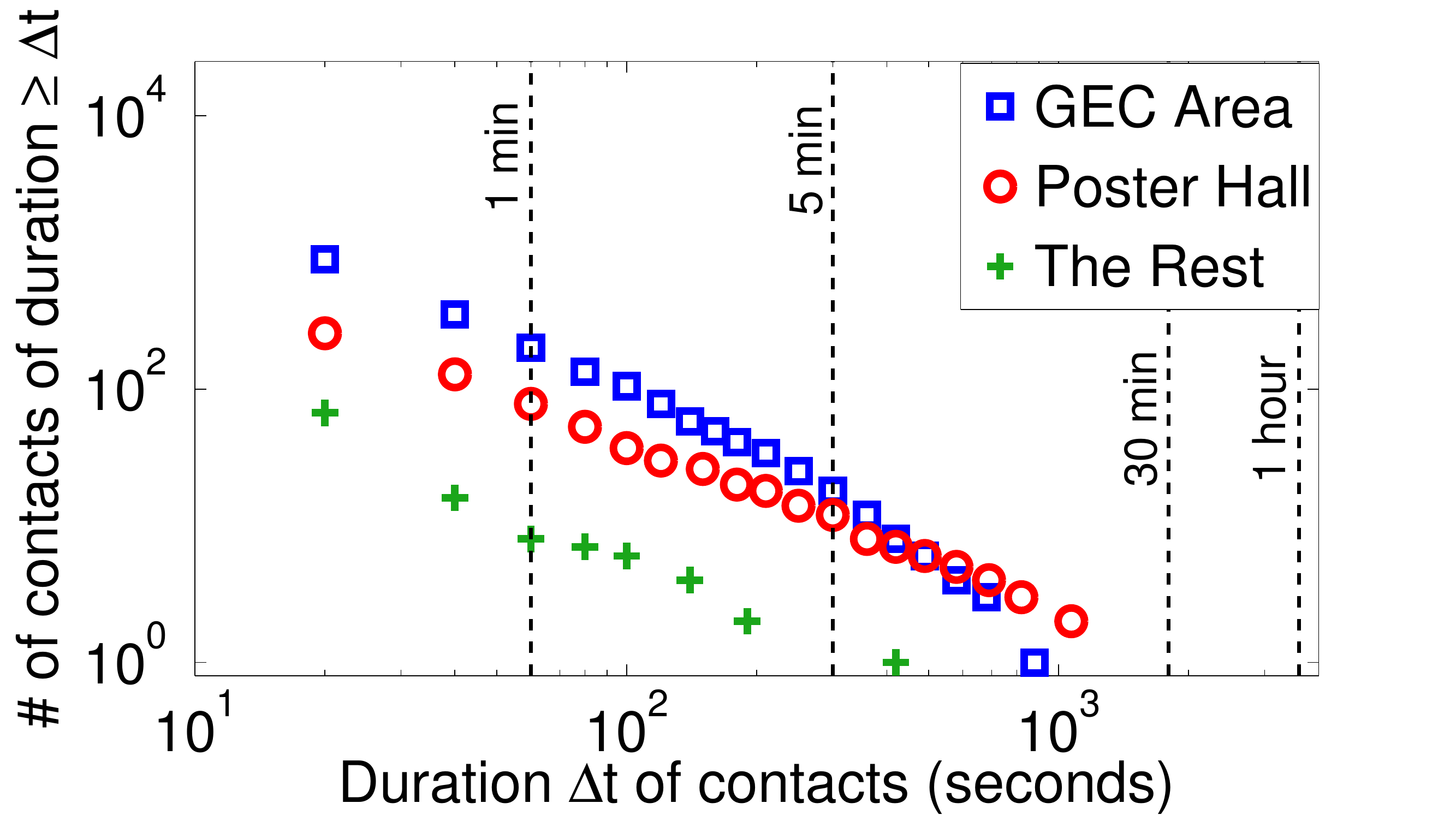} 
\end{minipage}
\end{center}

\caption{(Color online) Cumulative histograms of the durations of contacts in the 
three different areas within the SLC conference. Results for: (left) contacts within the DPP members, 
(middle) contacts between both conferences and (right) contacts within the GEC members.}
\label{contacts_fct_lieu} 
\end{figure}

To this aim, we show in Fig.~\ref{contacts_fct_lieu} the histograms of 
contact durations broken down
by category of contact and area. For the DPP contacts (left panel), 
the distributions measured in the various areas
have similar shapes, and the differences come from the overall number of contacts 
measured in each area
(as members of the DPP did not go much to the GEC area). On the other hand, for the 
contacts between both groups (middle panel) and for 
the GEC contacts (right panel), different slopes are observed depending on 
the area of interest. Broader
distributions are obtained in the Poster Hall, in particular for the contacts between 
GEC and DPP attendees:
the Poster Hall was therefore a more favorable setting for long cross-group 
contacts. This leads us to a somehow obvious remark: organizing activities 
in common physical spaces favors the
mixing between two groups.

\section{Co-located conferences case study: Cardinality constraint}
\label{sec:card}
We describe here the results of the test using only the cardinality constraint to test if the GEC
group behaves normally. We thus 
 consider the following simple Null Hypothesis: GEC behaves like any random group of $M=39$ individuals in the conference. 
For this first Null Hypothesis, the only constraint we impose to the bootstrap samples is therefore to have a cardinality equal to $M$. 

Applying the protocol described in \ref{subsec:protocol}, we first pick randomly with replacement $N_B=1000$ bootstraps 
samples of $39$ nodes.
For each sample, we compute the seven associated observables and normalize them as proposed in \ref{ssec:2C}.
\CHANGED{For each observable $Z$, the empirical distribution function $\hat{D}_z^b$
are computed from the bootstrap samples:  the $1-\alpha'$ two-sided acceptance interval
for $\hat{D}_z^b$ defines  what we call the ``normal behavior'' of a group under this constraint.}
We then obtain the divergences $d_{z}$ for each feature, and finally the triplet $(d,\chi^2,\sigma_u)$.


\begin{figure}

\begin{center}
 \begin{minipage}{0.48\linewidth}
 \begin{center}
 \includegraphics[width=0.8\textwidth]{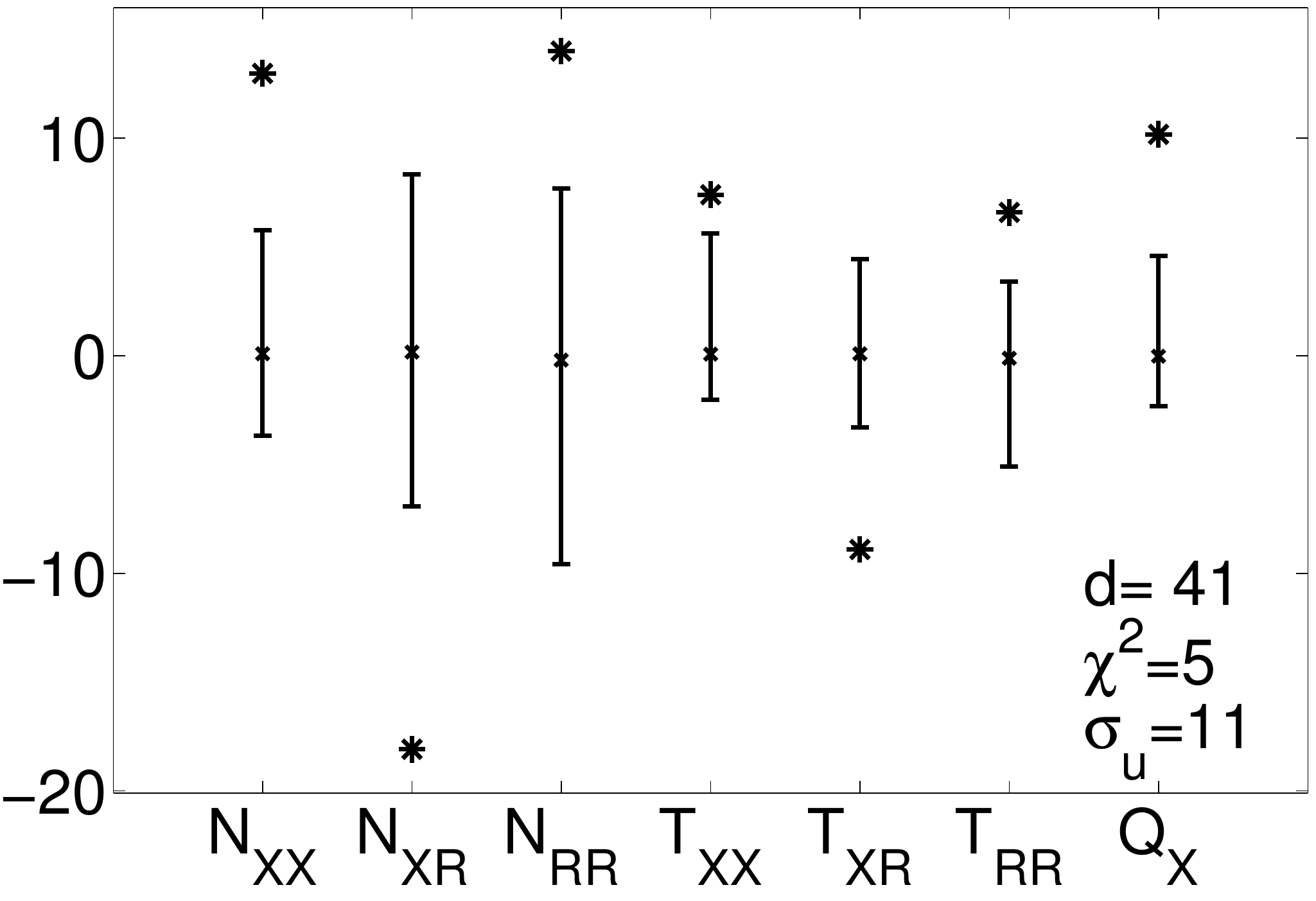} \\
 \vspace{0.5cm}
 \includegraphics[width=0.8\textwidth]{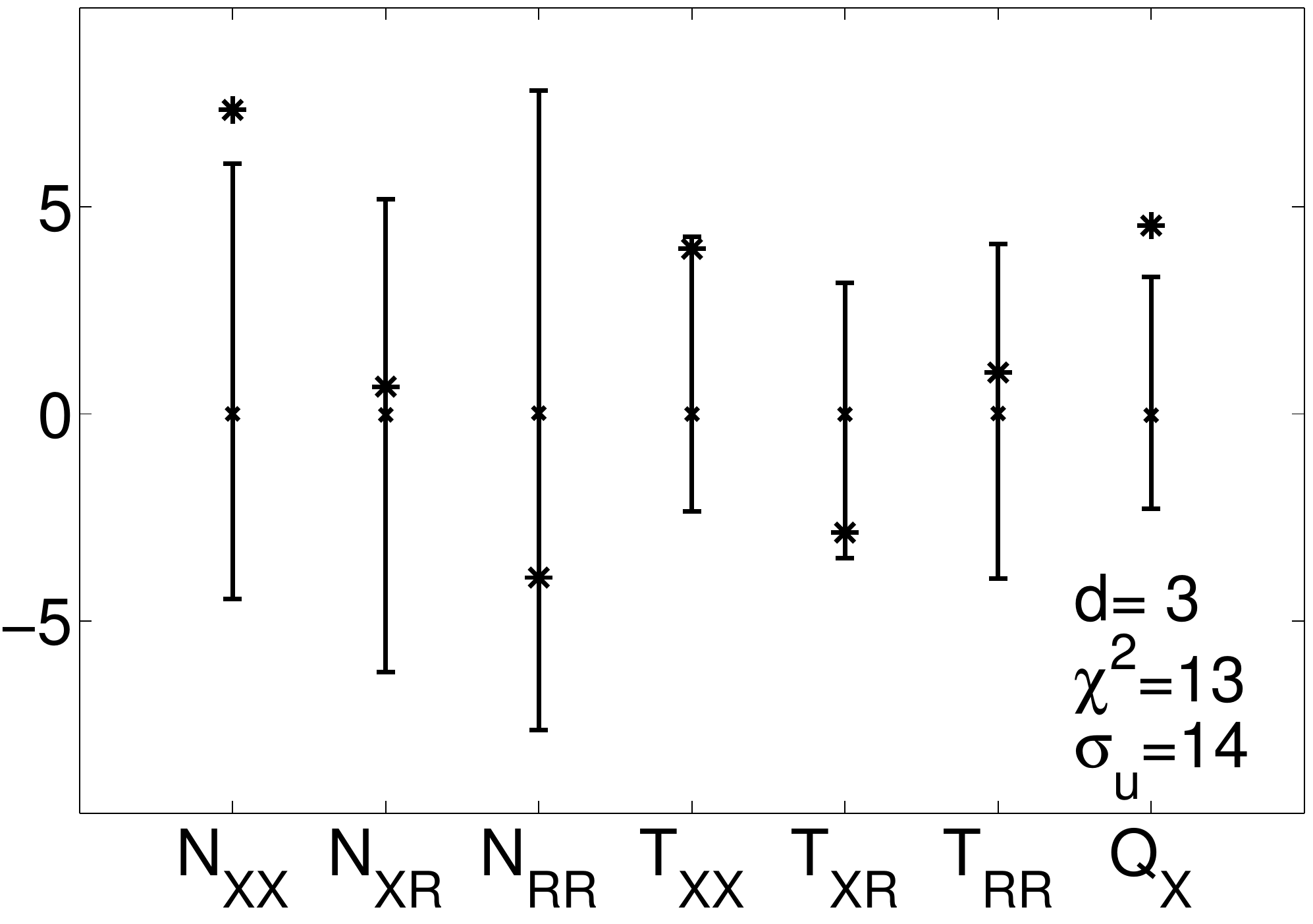}
 \end{center}
 \end{minipage}
\hfill
 \begin{minipage}{0.48\linewidth}
 \begin{center}
 \includegraphics[width=0.8\textwidth]{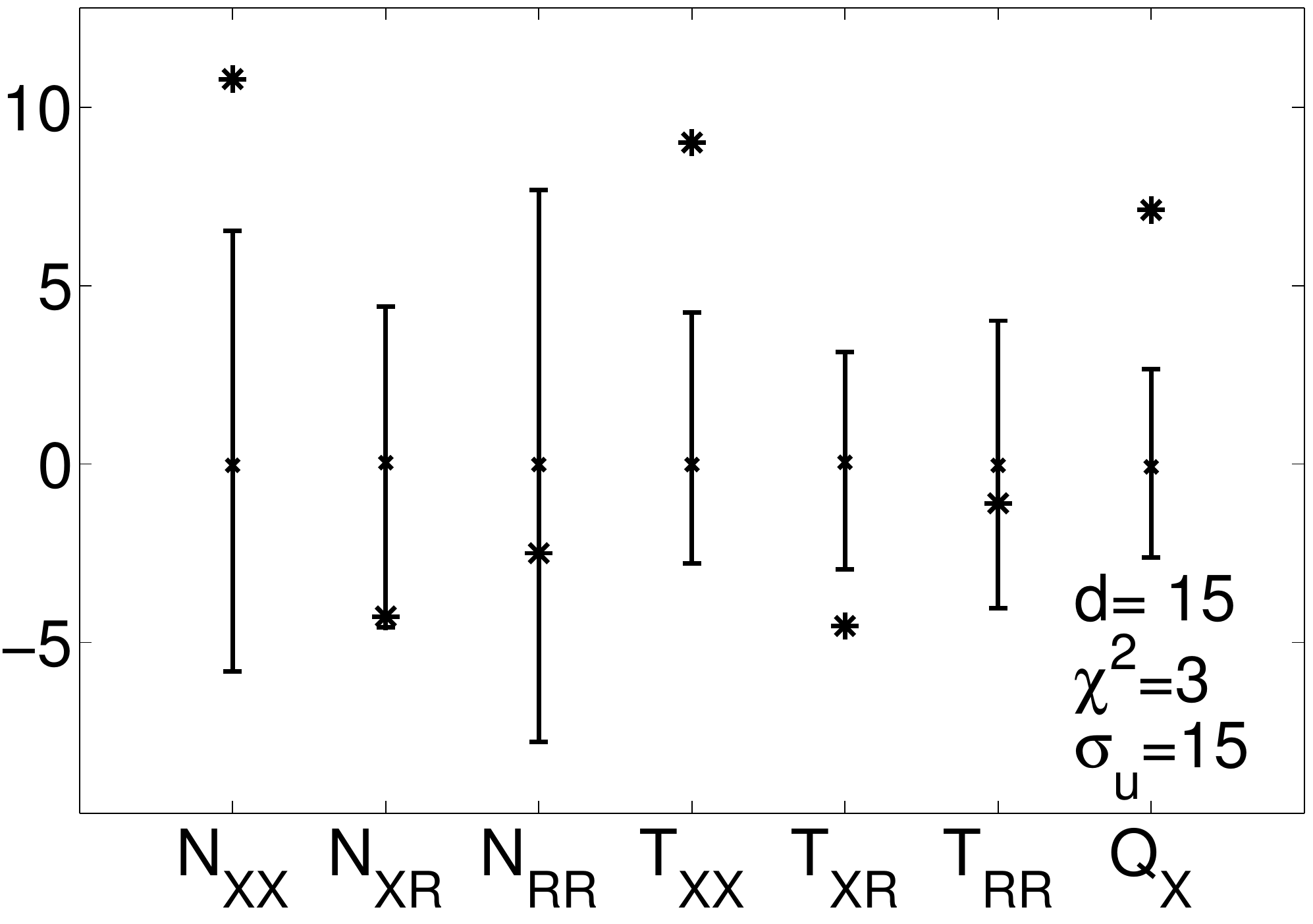}\\
  \vspace{0.5cm}
 \includegraphics[width=0.8\textwidth]{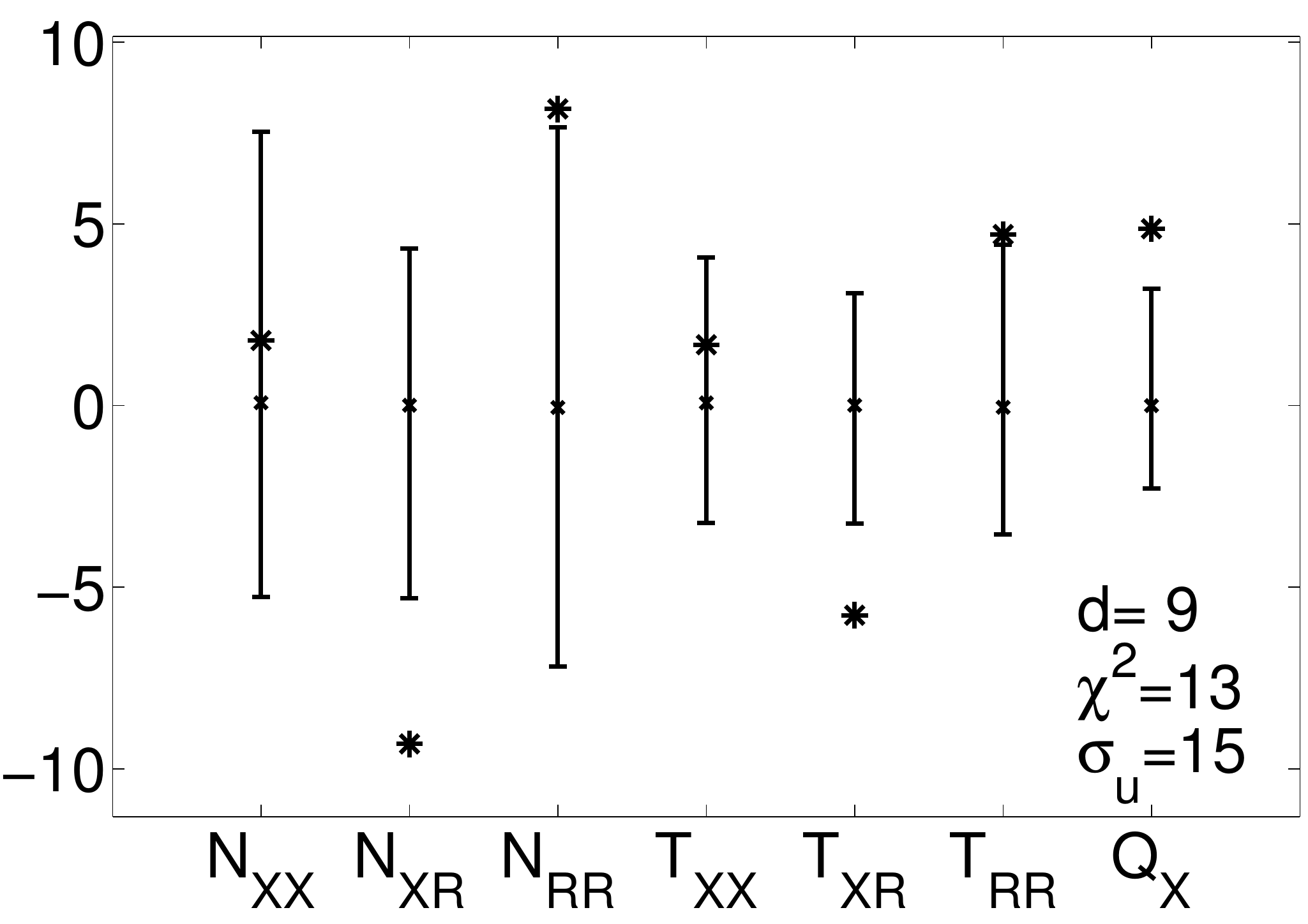}
  \end{center}
 \end{minipage}
\end{center}
\caption{
\CHANGED{Results of the test with same cardinality constraint for the four groups: GEC (top left), STP (top right), JUP (bottom left), and SEP (bottom right).
For each $z$, the two-sided acceptance interval with $1-\alpha'$ significance level 
are in black and the value $z^0$ is the black star. Here, $\alpha=5\%$, i.e.
 $\alpha'=\frac{\alpha}{F-f}=\frac{0.05}{7}=0.7\%$.
For each group $X^0=$ GEC, STP, JUP and SEP, the scalar $d$ (bottom right hand 
corner of each figure) is the total divergence between the acceptance interval 
of the bootstrap samples and the real data. $\chi^2$ and $\sigma_u$ are the 
two control parameters of the size of the bootstrap space.}
}
\label{Boxplot_nokeep} 
\end{figure}


\CHANGED{The top left plot of Figure~\ref{Boxplot_nokeep} summarizes the output
of this test for GEC: for each feature, the two-sided acceptance interval 
with $1-\alpha'$ significance level is shown by a black line
(its median being the black cross) for the $\hat{D}_z^b$ of the bootstrap samples, and the measured 
value of $z^0$ for GEC is figured by black stars. Finally, the values of $d$, $\chi^2$ and $\sigma_u$ 
are reported in the bottom right hand corner of the plot.} Figure~\ref{histos}a displays  
the two histograms yielding the two indicators $\sigma_u$ and $\chi^2$ in this case of cardinality constraint for $X^0$=GEC: 
they are both small. 
The other three plots of Figure~\ref{Boxplot_nokeep} show
the corresponding results for the three other groups (SEP, JUP, STP).

\CHANGED{The indicators $\chi^2$ and $\sigma_u$ are small enough for the four groups, indicating that
the bootstrap sets are large enough and that the test is fair according to the chosen Null Hypothesis, 
as commented in Section~\ref{subs:tradeoff} and Appendix~\ref{subs:CL_tradeoff}.
For all four groups, $d$ is non-null and the Null Hypothesis is rejected. }
In other words, none of these groups of individuals behaves similarly to a random group
of nodes with the same cardinality. These results do not come as a 
surprise since, as previously mentioned, these groups are somehow expected to behave as
communities and behave indeed as such: compared to the
bootstrap samples, they tend to have larger $Q_X$, $N_{XX}$, $N_{RR}$,
$T_{XX}$, $T_{RR}$ and smaller $N_{XR}$, $T_{XR}$.  Interestingly,
GEC's divergence is clearly larger than the others: this first test, even if somehow
na\"{\i}ve, hints at some difference between GEC and the other
groups.

\end{document}